\documentclass[%
 aip,
 jmp,%
 amsmath,amssymb,
preprint,%
author-numerical,
]{revtex4-1}
\newcommand{\be}{\begin{equation}}
\newcommand{\ee}{\end{equation}}

\newcommand{\ba}{\begin{array}}
\newcommand{\ea}{\end{array}}
\newcommand{\bea}{\begin{eqnarray}}
\newcommand{\eea}{\end{eqnarray}}
\newcommand{\nn}{\nonumber \\}
\newcommand{\qed}{\begin{flushright} $\square$
                  \end{flushright}
}
\newtheorem{theorem}{Theorem}

\newtheorem{corollary}{Corollary}

\usepackage{graphicx}
\usepackage{dcolumn}
\usepackage{bm}
\allowdisplaybreaks

\begin{document}


\title{Third-order superintegrable systems separable in parabolic coordinates}

\author{I. Popper}
\email{popperi@DMS.umontreal. ca}
\affiliation{%
Le D\'epartement de Math\'ematiques et de Statistique,
 Universit\'e de Montr\'eal. \\
 CP6128, Succursale Centre-Ville,  Montr\'eal (QC) H3C 3J7, Canada}

\author{S. Post}%
 \email{post@CRM.umontreal.ca}
\affiliation{ 
Centre de Recherches Math\'ematiques, Universit\'e de Montr\'eal.\\
 CP6128, Succursale Centre-Ville,  Montr\'eal (QC) H3C 3J7, Canada
}%

\author{P. Winternitz}
 \email{wintern@CRM.umontreal. ca}
\affiliation{%
Le D\'epartement de Math\'ematiques et de Statistique and Centre de Recherches Math\'ematiques,
Universit\'e de Montr\'eal.\\ CP6128, Succursale Centre-Ville, Montr\'eal   (QC) H3C 3J7, Canada 
}%

\begin{abstract}
In this paper, we investigate superintegrable systems which separate in parabolic coordinates and admit a third-order integral of motion. We give the corresponding determining equations and show that all such systems are multi-separable and so admit two second-order integrals. The third-order integral is their Lie or Poisson commutator.  We discuss how this situation is different from the Cartesian and polar cases where new potentials were discovered which are not multi-separable and which are expressed in terms of  Painlev\'e transcendents or  elliptic functions. 
\end{abstract}

\maketitle

\section{Introduction}
This article is part of a research program the aim of which is to identify all third-order superintegrable systems in two-dimensional Euclidean space. We recall that a superintegrable system is one that has more integrals of motion than degrees of freedom. We consider a classical or quantum Hamiltonian 
\be \label{1} H=\frac 12 (p_1^2+p_2^2)+V(x_1, x_2)\ee
with two integrals of motion
\be \label{2}X_a=\sum_{0\leq j+k\leq n}f_{a, jk}(x_1,x_2)p_1^jp_2, \qquad a=1,2,\ee
where $p_1, p_2$ are components of the momentum $\vec{p}$. In classical mechanics, the integrals $X_{1,2}$ Poisson commute with $H$, are well defined functions on phase space and the three functions $\{ H, X_1, X_2 \}$ are functionally independent. The functions $X_1$ and $X_2$ do not Poisson commute with each other; instead they generate a non-Abelian algebra, usually a polynomial one. In quantum mechanics, $H$ and $X_a$ are Hermitian operators in the enveloping algebra of the Heisenberg algebra (or some generalization of the enveloping algebra). The operators $X_a$ Lie commute with $H$, but not with each other. Instead of functional independence, we assume that $H$, $X_1$ and $X_2$  are algebraically independent. More specifically, we assume that no Jordan polynomial in the operators $H$, $X_1$ and $X_2$ vanishes. As indicated in (\ref{2}), we assume that $X_a$ are polynomials in the momentum. The "order of superintegrability" is the highest order of these polynomials. 

The best known superintegrable systems are the Kepler-Coulomb system\cite{Fock,Bargmann} with potential $V=\alpha/r$  and the harmonic oscillator\cite{JauchHill1940, MoshSmir} with $V=\alpha r^2$ . As a matter of fact, these are the only rotationally invariant superintegrable systems in $n$-dimensions ($n\geq 2$) (in agreement with Bertrand's theorem\cite{Bertrand, Goldstein} ). Both of these systems are quadratically superintegrable in that the Laplace-Runge-Lenz vector for $V=\alpha/r$ and the Fradkin (or quadropole) tensor for $V=\alpha r^2$ are both second-order in the momenta. 

Until recently, most studies of superintegrability concentrated on the second-order case\cite{FMSUW, WSUF, MSVW1967, Evans1990, SCQS}. At least in two-dimensional spaces, second-order superintegrable systems are well understood. All such systems in Euclidean spaces, spaces of constant curvature, and spaces of non-constant curvature with at least two second-order Killing tensors (Darboux spaces) have been classified \cite{KKMW, KKM20041, KKM20061, Eis}. 

Recently, infinite families of superintegrable systems with integrals of motion of arbitrary order have been discovered and investigated\cite{marquette2009super, TTW2009,TTW2010, PW20101, QuesneTTWodd, MPY2010,  KKM2010JPA, KKM2010quant, KKM2011Recurr, chanu2011polynomial, Ran2012}. 

A systematic search for third-order superintegrable systems was started in 2002, both in classical and quantum mechanics \cite{GW}. However, the first (to our knowledge), article on third-order integrals of motion was written considerably earlier by Drach \cite{Drach}. He considered the case of one third-order integral of motion (in addition to the Hamiltonian) in flat two-dimensional complex space in classical mechanics. He found 10 different complex potentials which allow a third-order integral. Later it was shown that 7 of them are actually quadratically superintegrable and the third-order integral is reducible, i.e. is the Poisson commutator of two second-order integrals \cite{Ran, Tsig1999}. 

The articles in Ref.'s \onlinecite{Gravel}, \onlinecite{GW} and \onlinecite{  TW20101} were devoted to superintegrable systems with one third-order and one first or second-order integral. A first-order integral in $E_2(\mathbb{R})$ exists only if the potential is translationally or rotationally invariant, i.e. $V=V(x)$ or $V=V(r).$  In the classical case, all such potentials are second-order superintegrable and hence known. In quantum mechanics, one class of new superintegrable potentials is obtained \cite{Gravel} and is expressed in terms of elliptical functions, e.g. 
\be V=\hbar^2\omega^2 sn^2(\omega x, k),\ee
where $\omega$ and $k$ are constants and $sn(\omega x, k)$ is a Jacobian elliptic function\cite{ByrdFriedman}. The existence of a second-order integral implies that $V(x_1, x_2)$ allows separation of variables in Cartesian, polar, parabolic or elliptic coordinates. Cartesian and polar coordinates were considered earlier \cite{Gravel, TW20101}. The study provided a number of new superintegrable potentials in the classical and, much more interestingly, in the quantum case. Indeed, quantum integrable systems with higher-order integrals of motion can be quite different from classical ones; a fact first noticed by Hietarinta \cite{Hiet1984, HG1989}. The case of quantum superintegrable systems is much richer than that of classical ones. Third-order superintegrability with separation of variables in the Schr\"odinger equation in Cartesian or polar coordinates lead to potentials expressed in terms of Painlev\'e transcendents ($P_I$, $P_{II}$ and $P_{IV}$ for Cartesian coordinates, $P_{VI}$ for polar ones). The potentials separable in Cartesian coordinates have been intensively studied in both classical and quantum mechanics \cite{MW2008, marquette2009painleve, marquette2009super}. 

The purpose of this article is to find all superintegrable systems that allow (at least) one third-order integral and a second-order integral that leads to separation in parabolic coordinates.

\section{The determining equations}
 Let us now assume that the Hamiltonian allows separation of variables in parabolic coordinates 
\[ x_1=\frac12 (\xi^2-\eta^2), \qquad x_2=\xi \eta.\]
The quantum mechanical Hamiltonian has the form 
\bea \label{H} H=-\frac{\hbar^2}{\xi^2+\eta^2}\left(\frac{\partial^2}{\partial \xi^2}+\frac{\partial^2}{\partial \eta^2}\right)+V(\xi,\eta),\\
\label{V} V(\xi, \eta)=\frac{W_1(\xi)+W_2(\eta)}{\xi^2+\eta^2}\eea
and there exists a second-order integral of the form 
\be Y=L_3p_2+p_2L_3+\frac{1}{\xi^2+\eta^2}\left(\xi^2W_2(\eta)-\eta^2W_1(\xi)\right) \ee
with 
\bea  p_1=-i\hbar \frac{\partial }{\partial x_1}=\frac{-i\hbar}{\xi^2+\eta^2}\left(\xi \frac{\partial}{\partial \xi}-\eta\frac{\partial}{\partial \eta}\right), \nn  p_2=-i\hbar\frac{\partial }{\partial x_2}=\frac{-i\hbar}{\xi^2+\eta^2}\left(\eta \frac{\partial}{\partial \xi}+\xi \frac{\partial}{\partial \eta}\right), \nn \nonumber L_3=-i\hbar(x_2p_1-x_1p_2)=\frac{i\hbar}{2}\left(\xi \frac{\partial }{\partial \eta}-\eta\frac{\partial }{\partial \xi}\right).\eea
A third-order integral will then have the form
\be \label{X} X=\sum_{j+k+\ell=3}A_{jk\ell}\left\{L_3^j, p_1^kp_2^\ell\right\} +\left\{g_1(x_1,x_2), p_1\right\} +\left\{ g_2(x_1, x_2), p_2\right\}. \ee
The brackets $\{\, , \, \}$ denote anti-commutators, $A_{jk\ell}$ are real constants and the functions $V$, $g_1$ $g_2$ obey the four partial differential equations presented in Ref. \onlinecite{GW} (in Cartesian coordinates). Here we need the equations in parabolic coordinates. To rewrite them in parabolic coordinates, it is convenient to replace the unknown functions $g_1(x_1,x_2)$ and $g_2(x_1, x_2)$ by $G_1(\xi, \eta)$, $G_2(\xi, \eta)$, putting 
\be\nonumber g_1(x_1, x_2)=\frac{\xi G_1(\xi, \eta)-\eta G_2(\xi,\eta)}{\xi^2+\eta^2}, \qquad g_2(x_1, x_2)=\frac{\eta G_1(\xi, \eta)+\xi G_2(\xi,\eta)}{\xi^2+\eta^2}.\ee
The four determining equations for the integral (\ref{X}), i.e. the commutativity condition $ [H, X]=0$, can be written as
\bea\label{Eq1}  G_{1, \xi}-{\frac {\xi G_{{1}} 
-\eta G_{{2}} }{{\xi}^{2}+{\eta}^{2}}}&=h_{
{1}} \left( \xi,\eta \right), 
\\\label{Eq2}
G_{{1, \eta}} +G_{{2, \xi}}-2\,{\frac { \eta G_{{1}}+\xi G_{{2}}  }{{\xi}^{2}+{\eta}^{2
}}} &=h_{{2
}} \left( \xi,\eta \right), \\
\label{Eq3}
G_{{2, \eta}} +{\frac {\xi G_{{1}} -\eta G_{{2}} }{{\xi}^{2}+{\eta}^{2
}}}&=h_
{{3}} \left( \xi,\eta \right), \eea
\bea \label{Eq4} G_1V_{\xi}+G_2V_{\eta}=\frac{\hbar^2}{4}\phi,\eea
with 
\begin{eqnarray*} h_1(\xi, \eta)&={\frac {3\,F_{{1}} V_{\xi} +F_{{2}} V_\eta }
{{\xi}^{2}+{\eta}^{2}}}
\\
h_2(\xi, \eta)&={\frac {2F_{{2}} V_{\xi} +2F_{{3}} V_\eta }
{{\xi}^{2}+{\eta}^{2}}}
\\
h_3(\xi,\eta)&={\frac {F_{{3}} V_\xi +3\,F_{{4}} V_\eta }{{\xi}^{2}
+{\eta}^{2}}}
,\end{eqnarray*}
and
\bea
  \phi &=&\frac{
F_1  V_{\xi\xi \xi}+F_2  V_{\xi\xi \eta}+F_3 V_{\xi\eta\eta}+F_4 V_{\eta\eta\eta}}{(\xi^2+\eta^2)^2} \\ &&
 +
 \frac{\left( 3\xi F_1+2\eta F_2-\xi F_3\right)V_{\xi\xi} +\left(\eta F_2-\xi F_2-\eta F_3+\xi F_4\right)V_{\xi, \eta}+
\left(\eta F_2-2\xi F_3-3\eta F_4\right)V_{\eta\eta}}{(\xi^2+\eta^2)^3}\nn&&
-\left( \frac{
3(\xi^2-\eta^2)(F_1-F_3)+6\xi\eta (F_2-F_4)}{(\xi^2+\eta^2)^4}
-4A_{300}\eta(\xi^2+\eta^2)+2A_{201}(\xi+\eta)\right)V_\xi\nn  &&
-\left(\frac{ 6\xi\eta (F_1-F_3)-3(\xi^2-\eta^2)(F_2-F_4)}{(\xi^2+\eta^2)^4} -4A_{300}\xi(\xi^2+\eta^2)-2A_{201}(\xi+\eta)\right)V_\eta.\nonumber\eea
The subscripts denote partial derivatives and the expressions $F_1, F_2, F_3, F_4$ are polynomials in $\xi$ and $\eta$:{\small 
 \bea\nonumber F_1 &=&{\eta}^{3}A_{{003}}+\xi\,{\eta}^{2}A_{{012}}+{\xi}^{3}A_{{030}}-\frac{1}{2}\,{\eta
}^{3} \left( {\xi}^{2}+{\eta}^{2} \right) A_{{102}}-\frac{1}{2}\,\xi\,{\eta}^{
2} \left( {\xi}^{2}+{\eta}^{2} \right) A_{{111}}-\frac{1}{2}\,\eta\,{\xi}^{2}
 \left( {\xi}^{2}+{\eta}^{2} \right) A_{{120}}\\  &&\nonumber
 +\frac{1}{4}\,{\eta}^{3} \left( {\xi}^{2}+{\eta}^{2} \right) ^{2}A_{{201}}+\frac{1}{4}\,\xi\,{\eta}^{2}
 \left( {\xi}^{2}+{\eta}^{2} \right) ^{2}A_{{210}}-\frac{1}{8}\,{\eta}^{3}
 \left( {\xi}^{2}+{\eta}^{2} \right) ^{3}A_{{300}}+{\xi}^{2}\eta\,A_{{021}}\eea
  \bea\nonumber F_2&=&3\,\xi\,{\eta}^{2}A_{{003}}-\eta\, \left( {\eta}^{2}-2\,{\xi}^{2}
 \right) A_{{012}}-3\,{\xi}^{2}\eta\,A_{{030}}-\frac{1}{2}\,\xi\,{\eta}^{2}
 \left( {\xi}^{2}+{\eta}^{2} \right) A_{{102}}+\frac{1}{2}\,{\eta}^{3} \left( 
{\xi}^{2}+{\eta}^{2} \right) A_{{111}}\\\nonumber  &&\nonumber
+\frac{1}{2}\,\xi\, \left( {\xi}^{2}+2\,
{\eta}^{2} \right)  \left( {\xi}^{2}+{\eta}^{2} \right) A_{{120}}-\frac{1}{4}
\,\xi\,{\eta}^{2} \left( {\xi}^{2}+{\eta}^{2} \right) ^{2}A_{{201}}-\frac{1}{4}\eta\, \left( {\eta}^{2}+2\,{\xi}^{2} \right)  \left( {\xi}^{2}+{\eta}^{2} \right) ^{2}A_{{210}}\\  &&\nonumber
+\frac{3}{8}\,\xi\,{\eta}^{2} \left( {\xi}^{2}+{\eta}^{2} \right) ^{3}A_{{300}}
-A_{{021}}\xi\, \left( 2\,{\eta}^{2}-{\xi}^{2} \right)\eea
 \bea\nonumber F_3&=&3\,{\xi}^{2}\eta\,A_{{003}}-\xi\, \left( 2\,{\eta}^{2}-{\xi}^{2}
 \right) A_{{012}}+3\,\xi\,{\eta}^{2}A_{{030}}+\frac{1}{2}\,\eta\,{\xi}^{2}
 \left( {\xi}^{2}+{\eta}^{2} \right) A_{{102}}+\frac{1}{2}\,{\xi}^{3} \left( {
\xi}^{2}+{\eta}^{2} \right) A_{{111}}\\\nonumber  &&\nonumber
-\frac{1}{2}\,\eta\, \left( {\eta}^{2}+2\,{\xi}^{2} \right)  \left( {\xi}^{2}+{\eta}^{2} \right) A_{{120}}-\frac{1}{4}
\,\eta\,{\xi}^{2} \left( {\xi}^{2}+{\eta}^{2} \right) ^{2}A_{{201}}+\frac{1}{4}\,\xi\, \left( {\xi}^{2}+2\,{\eta}^{2} \right)  \left( {\xi}^{2}+{\eta}^{2} \right) ^{2}A_{{210}}\\  &&\nonumber
-\frac{3}{8}\,\eta\,{\xi}^{2} \left( {\xi}^{2}+{\eta}^{2} \right) ^{3}A_{{300}}+
\eta\,A_{{021}} \left( {\eta}^{2}-2\,{\xi}^{2} \right)\eea
 \bea\nonumber F_4&=&{\xi}^{3}A_{{003}}-{\xi}^{2}\eta\,A_{{012}}-{\eta}^{3}A_{{030}}+\frac{1}{2}\,{\xi}
^{3} \left( {\xi}^{2}+{\eta}^{2} \right) A_{{102}}-\frac{1}{2}\,\eta\,{\xi}^{2
} \left( {\xi}^{2}+{\eta}^{2} \right) A_{{111}}
+\frac{1}{2}\,\xi\,{\eta}^{2}\left( {\xi}^{2}+{\eta}^{2} \right) A_{{120}}\\\nonumber &&
+
\frac{1}{4}\,{\xi}^{3} \left( {\xi}^{2}+{\eta}^{2} \right) ^{2}A_{{201}}-
\frac{1}{4}\,\eta\,{\xi}^{2} \left( {\xi}^{2}+{\eta}^{2} \right) ^{2}A_{{210}}+
\frac{1}{8}\,{\xi}^{3} \left( {\xi}^{2}+{\eta}^{2} \right) ^{3}A_{{300}}
+\xi\,{\eta}^{2}A_{{021}}.\eea}

The determining equations (\ref{Eq1}-\ref{Eq3}) are the same in  classical and quantum mechanics but (\ref{Eq4}) contains the Planck constant on the right hand side. The corresponding equation in classical mechanics is obtained by taking the limit $\hbar\rightarrow 0$ so (\ref{Eq4}) is greatly simplified. Hence the difference between classical and quantum integrability (and superintegrability) for third-order integrals of motion. 

The system (\ref{Eq1}-\ref{Eq4}) is overdetermined. The first three equations imply a linear compatibility condition for the potential
\bea \label{cc} \ba{rl}0=& F_3 V_{{\xi \xi \xi}} +(3\,F_{{4}}-2\,F_{{2}})V_{{\xi \xi \eta}} +(3\,F_{{1}}-2\,F_{{3}})V_{{\xi \eta \eta}}+F_{{2}}V_{{\eta \eta \eta}}\\&+
\left(2(F_{{3\,\xi}}-F_{{2\,\eta}}) -{\frac {3\xi F_{{1}} 
 -6\eta F_{{2}}  
+7\,\xi\,F_{{3}} }{  {\xi}^{2}+{\eta}^{2} }}\right)V_{{\xi \xi}}+\left(2(F_{{2\,\eta}}-F_{{3\,\xi}}) -{\frac  {3\,\eta\,F_{{4}}  -6\,\xi\,F_{{3}}+7\eta F_{{2}}   }{  {\xi}^{2}+{\eta}^{2} }}\right)V_{{\eta\eta }}\\&+
\left(2\left(3{F_{{1\,\eta}}}- {F_{{2\,\xi}}
}- {F_{{3\,\eta}}}+3 {F_{{4\,\xi}}} \right)-{\frac {21\,\eta\,F_
{{1}}  -5\,\eta\,F_{{3}} 
  -5\,\xi\,F_{{2}}  +21\,\xi\,F_{{4}}}{ {\xi}^{2}+{\eta}^{2} }}\right)V_{{\xi \eta}}\\&+
 A\,V_\eta + B\,V_\xi ,\ea\eea
  where
 \bea\nonumber A&=& {F_{{2\,\eta\,\eta}}}-2{F_{{3\,
\eta\,\xi}}}+3{F_{{4\,\xi\,\xi}}}
+{\frac {-7\eta\,F_{{2\,\eta}} -\xi\,F_{{2\,\xi}}+6\xi\,F_{{3\,\eta}} +6\eta\,F_{{3\,\xi}}-3\eta\,F_{{4\,\eta}}-21\xi\,F_{{4\,\xi}}  }{{\xi}^{2}+{\eta
}^{2} }}  \nn &&
+2\,{\frac {21\,{\xi}^{2}F_
{{4}}  +F_{{2}}  {\xi}^{
2}+7\,{\eta}^{2}F_{{2}}  -12\,\xi\,\eta\,F_{{3}
}  +3\,F_{{4}}  {\eta}^{
2}}{ \left( {\xi}^{2}+{\eta}^{2} \right) ^{2}}} \nonumber
\eea
\bea \nonumber  B&=& 3 F_{{1\,\eta\,\eta}}-2 {F_{{2\,
\eta\,\xi}}}+{F_{{3\,\xi,\xi}}}-{\frac {21\eta\,F_{{1\,\eta}}+3\xi\,F_{{1\,\xi}}-6\xi\,F_{{2\,\eta}} -6\eta\,F_{{2\,\xi}}+\eta\,F_{{3\,\eta}}+7\xi\,F_{{3\,\xi}
} }{  {\xi}^{2}+{\eta}^
{2} }}\nn &&+2\,{\frac {F_{{3}}
  {\eta}^{2}+3\,F_{{1}}  {\xi}^{2}+21\,{\eta}^{2}F_{{1}}  -12\,
\xi\,\eta\,F_{{2}}  +7\,{\xi}^{2}F_{{3}}
  }{ \left( {\xi}^{2}+{\eta}^{2} \right) ^{2}}}. \nonumber\eea

 Compatibility between the first three determining equations (\ref{Eq1}-\ref{Eq3}) and the fourth one (\ref{Eq4}) requires three more conditions, this time nonlinear ones. Indeed, solving (\ref{Eq4}) for $G_2$, we have 
\be\label{G2} G_2=\frac{1}{V_{\eta}}\left(\frac{\hbar^2}{4}\Phi -V_\xi G_1\right), \qquad V_\eta\ne 0.\ee
Replacing $G_2$ from (\ref{G2}) into  (\ref{Eq1}-\ref{Eq3}), the system can then be solved for $G_1,$
\be\label{G12}  G_1=\frac{h_4}{h_5}, \qquad G_2=\frac{h_5\hbar^2 \Phi-4V_\xi h_4}{4 h_5 V_\eta}, \qquad h_5\ne 0,\ee
with
\begin{eqnarray*}
 h_4&=&(\xi^2+\eta^2)\left(4h_3V_\eta^3+4h_2V_\xi V_\eta^2+4h_1 V_\xi^2V_\eta-\hbar^2 V_\eta^2\Phi_{ \eta}-\hbar^2 V_\xi V_\eta \Phi_{ \xi}-\hbar^2\left(V_{\eta \eta} V_\eta-V_{\xi\eta} V_\xi \right)\Phi\right)\nn
 &&
+4\eta V_\xi^2-4\eta V_\eta^2-8\xi V_\eta V_\xi\nn
 h_5&=&4(\xi^2+\eta^2)\left(V_\xi V_\eta (V_{\xi \xi}-V_{\eta\eta}+(V_\eta^2-V_\xi^2)V_{\xi\eta}\right) +4(\eta V_\xi-\xi V_\eta)(V_\xi^2+V_\eta^2).\end{eqnarray*}
Replacing (\ref{G12}), into (\ref{Eq1}-\ref{Eq3}), gives the three additional non-linear compatibility conditions on the potential, namely
\bea\label{ccnl1} \left( \frac{h_4}{h_5}\right)_\xi +\frac{\eta\hbar^2h_5 \Phi +4h_4\left(\xi V_\eta+\eta V_\xi\right)}{4 h_5V_\eta(\xi^2+\eta^2)}=h_1\\ \label{ccnl2}
\left( \frac{h_4}{h_5}\right)_\eta +\left(\frac{\hbar^2 h_5\Phi-4V_\xi h_4}{4 h_5 V_\eta}\right)_\xi+\frac{\xi\hbar^2 h_5\Phi +4h_4\left(\eta V_\eta-\xi V_\xi\right)}{2h_5V_\eta(\xi^2+\eta^2)}=h_2\\
\left(\frac{\hbar^2 h_5\Phi-4V_\xi h_4}{4h_5 V_\eta}\right)_\eta -\frac{\eta\hbar^2h_5\Phi+4h_4\left(\xi V_\eta+\eta V_\xi\right)}{4h_5V_\eta(\xi^2+\eta^2)}=h_3\label{ccnl3}
.\eea

\section{General forms for $W_1$ and $W_2$}\label{general}
In order to determine all possible potentials which separate in parabolic coordinates and admit a third-order integral of motion, we begin with the linear compatibility condition (\ref{cc}). Replacing $V$ with the form as in (\ref{V}), the compatibility condition can be differentiated to obtain a system of linear ordinary differential equations  (ODEs) for $W_1$ and $W_2.$ These equations are given in Appendix \ref{eqs} for $W_1$. Notice that  interchanging $\xi$ and $\eta$ has the effect of changing the sign of the coefficients to $(-1)^{j+k}A_{jk\ell}.$ Thus, equations for $W_2$ are of the same form as (\ref{eq1}-\ref{eq11}), up to a change in sign in these constants. We begin our search for the admissible potentials by solving equations (\ref{eq1}-\ref{eq11}). 

\begin{theorem}\label{thmW1} Given a Hamiltonian which admits a third-order integral of motion  with a potential which separates in parabolic coordinates as in (\ref{V}). Then any admissible terms in the potential are included in
\bea\label{W1}  W_1(\xi)=\sum_{j=0}^{16}c_j\xi^j+c_{-2}\xi^{-2}+c_{-4}\xi^{-4}+c_{-6}\xi^{-6}+\frac{\alpha_1\xi \ln \left(\xi+\sqrt{\xi^2+B_1}\right)+\alpha_2\xi}{\sqrt{\xi^2+B_1}},\\
\label{W2}  W_2(\eta)=\sum_{j=1}^{16}d_j\eta^j+d_{-2}\eta^{-2}+d_{-4}\eta^{-4}+d_{-6}\eta^{-6}+  \frac{\beta_1 \ln \left(\eta+\sqrt{\eta^2+B_2}\right)+\beta_2\eta}{\sqrt{\eta^2+B_2}}.\eea
\end{theorem}
In particular, any solution of the system (\ref{cc}) has the form (\ref{V}) with $W_1$ and $W_2$ as in  (\ref{W1}) and (\ref{W2}) with appropriately chosen constants. 

{\bf Proof:}
Beginning with (\ref{eq1}), in the case that $A_{300}\ne0,$
the solutions are given by
\be \label{eq1solns}
W_{1}=\sum_{i=0}^{14} c_i\xi^i +c_{-2}\xi^{-2}+c_{-4}\xi^{-4}.\ee
Similarly, for (\ref{eq2}), in the case that $A_{210}\ne0$,  the solutions are 
\be \label{eq2solns}
W_{1}=\sum_{i=0}^{15} c_i\xi^i +c_{-2}\xi^{-2}.\ee
Equation (\ref{eq3}) in addition requires that $c_{14}=c_{15}=0.$

Now assume $A_{300}=A_{210}=0$. This sets (\ref{eq1}-\ref{eq3}) to be satisfied identically and (\ref{eq4}) becomes
\be \label{eq4s} - \left(\left( {\xi}^{2}A_{{201}}+2\,A_{{120}} \right) {\frac {d^{17}}{d{\xi
}^{17}}} +33 \xi\,A_{{201}} {\frac {d^{16}}{d{\xi}
^{16}}}+255A_{{201}}
 {\frac {d^{15}}{d{\xi}^{15}}}\right)W_{{1}} 
  =0.\ee
 The solutions of (\ref{eq4s}), assuming $A_{201}\ne 0,$ are 
 \be \label{eq4ssolns1} W_1=\sum_{i=0}^{14} c_{i}\xi^i +\frac{\alpha_1\xi \ln \left(\xi+\sqrt{\xi^2+B_1}\right)+\alpha_2\xi}{\sqrt{\xi^2+B_1}}.\ee
 with $A_{120}= B_1 A_{201}/2.$
If $A_{201}=0$, the solutions are 
\be \label{eq4ssolns2} W_{1}=\sum_{i=0}^{16} c_i\xi^i .\ee
Now assume $A_{300}=A_{210}=A_{201}=A_{120}=0,$ (\ref{eq1}-\ref{eq4}) are then identically satisfied and  (\ref{eq5}) becomes
\be \label{eq5s}  \frac12\left(\left(A_{{102}}{\xi}^{2} +2A_{{021}}\right) {\frac {d^{17}}{d{\xi
}^{17}}} +{ {255}} A_{{102}}{\frac {
d^{15}}{d{\xi}^{15}}}+{
 {33}} \xi\,A_{{102}} {\frac {d^{16}}{d{\xi}^{16}}}\right)W_{{1}}=0.\ee
The solutions of (\ref{eq5s}) for $A_{102}\ne 0$ are as (\ref{eq4ssolns1}) with $A_{021}=2A_{102}B_{1}$ and for $A_{102}=0$ are as (\ref{eq4ssolns2}). 

If we now assume $A_{300}=A_{210}=A_{201}=A_{120}=A_{102}=A_{021}=0,$ 
(\ref{eq6}) becomes
\be \label{eq6s}  A_{003}\left[3\, {\xi}^{2}{\frac {d^{17}}{d{\xi}^{17}}}+111\xi  {\frac {d^{16}}{d{\xi}^{16}}}+969 {\frac {d^
{15}}{d{\xi}^{15}}}\right]W_{{1}} =0,\ee
the solutions of which are given by (\ref{eq1solns}).

Now, assume $A_{300}=A_{210}=A_{201}=A_{120}=A_{102}=A_{021}=A_{003}=0,$ this implies (\ref{eq1}- \ref{eq6}) as well as (\ref{eq9}-\ref{eq8}) are identically satisfied. Equation (\ref{eq10}) becomes
\bea\label{eq10s}   \frac{\xi\left( A_{{111}}{\xi}^{2}+6\,A_{{030}}-4\,A_{{012}} \right)}{2}
\frac {d^{17}}{d{\xi}^{17}}W_1+ \left({\frac {51}{2}}\,A_{{111}}{\xi}^{2}  -36\,A
_{{12}}+54\,A_{{030}}\right) {
\frac {d^{16}}{d{\xi}^{16}}}W_{{1}} \nn +{\frac {813\xi A_{111}}{2
}} {\frac {d^{15}}{d{\xi}^{15}}}W_{{1}}   +2016A_{111} {\frac {d^{14}}{d{\xi}^{14}}}W_{
{1}} =0.\eea
The general solutions of (\ref{eq10s}) are of the form
\be \label{eq10ssolns1} W_1=\sum_{i=0}^{13} c_{i}\xi^i +c_{-2}\xi^{-2}+\frac{\alpha_1\xi \ln \left(\xi+\sqrt{\xi^2+B_1}\right)+\alpha_2\xi}{\sqrt{\xi^2+B_1}},\ee
 with $A_{030}=\frac16 B_{1} A_{111}+\frac23 A_{012}$ for $A_{111}\ne 0$ and of the form (\ref{eq2solns}) for $A_{111}=0$. 

Now assume that $A_{300}=A_{210}=A_{201}=A_{120}=A_{102}=A_{021}=A_{003}=A_{111}= 3A_{030}-2A_{012}=0,$ this implies that (\ref{eq11}) has the form 
\be \label{eq11s}    A_{012}\left( {\xi}^{3}{\frac {d^{17}}{d{\xi}^{17}}} +57{\xi}^{2}{\frac {d^{16}}{d{\xi}^{16}}}+1023\xi {\frac {d^{
15}}{d{\xi}^{15}}}+5760 {\frac {d^{14}}{d{\xi}^{14}}}
 \right)W_{{1}} 
 =0. \ee
 The solutions of (\ref{eq11s}) for $ A_{012}\ne 0$ are given by 
 \be \label{eq11ssolns} W_1=\sum_{i=0}^{14} c_i\xi^i +c_{-2}\xi^{-2}+c_{-4}\xi^{-4}+c_{-6}\xi^{-6}.\ee
This is the final case, since assuming $A_{012}=0$ in addition to  $A_{300}=A_{210}=A_{201}=A_{120}=A_{102}=A_{021}=A_{003}=A_{111}= 3A_{030}-2A_{012}=0$ has all the A's equal 0 and so there is no longer a third order term in our constant of motion. 

Thus, the most general form of the function $W_{1}$ is given by (\ref{W1}). By direct analogy, the most general form of the function $W_2$ is given by (\ref{W2}). \qed
\begin{corollary} The potential for any 3rd-order superintegrable system which separates in parabolic coordinates satisfies a non-trivial system of linear ODEs for both $W_1$ and $W_2$. 
\end{corollary}
{\bf Proof }
As shown above, the compatibility conditions (\ref{cc}) are satisfied identically if and only if all the $A_{ijk}$ are 0. On the other hand, if there is a non-zero $A_{ijk}$ then the functions $W_1$ and $W_2$ in the potential will satisfy some linear ODEs. \qed 

\section{The absence of irrational terms}
In this section, we show that the only possible form of the potential is as a rational function of $\xi$ and $\eta.$ We shall show this by contradiction. Namely, we consider the case that  either $\alpha_1\ne 0$ or both  $\alpha_2\ne0$ and  $B_1 \ne 0$ in (\ref{W1}).  

If we substitute the general form (\ref{W1}) into equations (\ref{eq1}-\ref{eq11}), it is immediate from the previous section that $A_{300}=A_{210}=0$.  We also obtain the following possible restrictions on the constants, as suggested in the previous section:
\be \label{A1} \left\{\ba{c} A_{{003}}=A_{021}=\frac{B_1(A_{102}-A_{120})}{2} ,\\
 A_{{012}}=A_{{030}}=\frac{A_{{111}}B_{{1}}}{2},\\ A_{{201}}={\frac {2A_{{120}}}{B_{{1}}}}.\ea\right\}\ee
In the case that $B_1\ne 0,$ $W_1$ becomes
\be\label{W1m2} W_1=c_{{0}}+c_{{1}}\xi+c_{{2}}{\xi}^{2}+c_{{3}}{\xi}^{3}+c_{{4}}{\xi}^{4}+
c_{{5}}{\xi}^{5}+c_{{6}}{\xi}^{6}+{\frac {c_{{-2}}}{{\xi}^{2}}}+\frac{\alpha_1\xi \ln \left(\xi+\sqrt{\xi^2+B_1}\right)+\alpha_2\xi}{\sqrt{\xi^2+B_1}},
\ee
with the following cases 
\bea \label{casesm2}\ba{l} 
\{B_1\ne 0, c_{-2}=c_5=c_6=0\}\\
\{B_1\ne 0, A_{201}=0, c_{-2}=c_6=0\}\\
\{B_1\ne 0, A_{111}=A_{201}=0, c_{-2}=0\}\\
\{B_1\ne 0, A_{102}=A_{201}=0. \}\ea \eea
In the case that $\alpha_1 \ne 0$ and $B_1 =0$, $W_1$ becomes 
\be \label{W1m4} W_1=c_{{0}}+c_{{1}}\xi+c_{{2}}{\xi}^{2}+c_{{3}}{\xi}^{3}+c_{{4}}{\xi}^{4}+
c_{{5}}{\xi}^{5}+c_{{6}}{\xi}^{6}+{\frac {c_{{-2}}}{{\xi}^{2}}}+{\frac {c_{{-4}}}{{\xi}^{2}}}+
\alpha_{{1}}\ln (\xi),\ee
with the following cases 
\bea \label{casesm4}\ba{l} 
\{\alpha_1\ne 0, B_1=0, c_{-4}=c_5=c_6=0\}\\
\{\alpha_1\ne 0, B_1=0, A_{201}=0, c_{-4}=c_6=0\}\\
\{\alpha_1\ne 0, B_1=0, A_{111}=A_{201}=0, c_{-4}=0\}\\
\{\alpha_1\ne 0, B_1=0, A_{102}=A_{201}=0 \}.\ea \eea
Note that  when $B_1=0$,  $\alpha_2$ becomes an additive constant which is absorbed into $c_0$.

Next, we substitute these cases (\ref{casesm2}) or (\ref{casesm4}),  along with the forms of $W_1$ (\ref{W1m2}) or (\ref{W1m4}) and $W_2$ (\ref{W2}), into the compatibility condition (\ref{cc})  to obtain
\be0=M_{2}ln\left( \eta^2+\sqrt{\eta^2+B_2}\right)+M_{1} \sqrt{\eta^2+B_2} +M_{0}\ee
where the $M_i$ are rational functions which are too long to be presented here but are available from the authors upon request. 
Solving $M_{2}=0$ gives 3  possibilities: either $\beta_1=0$, $B_2=-B_1$ or, in the case that $B_1$ is not zero, $B_2$ may vanish if additionally $A_{111}=0$ and $ A_{102}=B_1A_{201}/2$. 

Solving $M_1=0$ and $M_0=0$ gives similar forms for $W_2:$ If $B_1\ne 0$, then $W_2$ has the form 
\be\label{W2m2} W_2=d_1\eta+c_{{2}}{\eta}^{2}-c_{{4}}{\eta}^{4}+c_{{6}}{\eta}^{6}+{\frac {c_{{-2}}}{{\eta}^{2}}}+\frac{-\alpha_1\eta \ln \left(\eta+\sqrt{\eta^2+B_2}\right)+\beta_2\eta}{\sqrt{\eta^2+B_2}}
\ee
and if $B_1=0$, then it has the form 
\be \label{W2m4} W_2=c_{{2}}{\eta}^{2}-c_{{4}}{\eta}^{4}+c_{{6}}{\eta}^{6}+{\frac {c_{{-2}}}{{\eta}^{2}}}-\frac{c_{-4}}{\eta^{4}}-\alpha_1 \ln(\eta).\ee
In both cases, $c_3$ and $c_5$ are also required to be 0 as well and  as several additional sets of constraints which are required to completely solve (\ref{cc}). We shall return to some of these cases later. 

To obtain a contradiction for the potentials admitting logarithmic singularities, we now turn our attention to the non-linear compatibility conditions. Beginning with Eq. (\ref{ccnl1}), we clear the denominator and consider the equation
\be\label{CCN2} 0=h_5^2V_\eta(\xi^2+\eta^2)\left[\left( \frac{h_4}{h_5}\right)_\xi +\frac{\eta\hbar ^2h_5\Phi +4h_4\left(\xi V_\eta+\eta V_\xi\right)}{4h_5V_\eta(\xi^2+\eta^2)}-h_1\right]. \ee
In the case that both $\alpha_1$ and $B_1$ are assumed non-zero, equation (\ref{CCN2}) will have polynomial dependence on the quantity $ln\left( \xi^2+\sqrt{\xi^2+B_1}\right)$.  Substituting the forms of the potential obtained above (\ref{W1m2}) and (\ref{W2m2}) into this quantity,(\ref{CCN2}), gives 
\bea 0&=&32\alpha^6B_1^3 K ln^6\left( \xi^2+\sqrt{\xi^2+B_1}\right)+\mathcal{O}\left(ln^5\left( \xi^2+\sqrt{\xi^2+B_1}\right)\right),\\
 K&=&9 A_{111}\eta\xi^{14}+ \frac{9A_{111}}{2}(16\eta^2+5B_1)\xi^{12}\nn &&
+\left(2(48A_{102} -23A_{201})\eta^2-B_1(16A_{102}-11A_{201}) \right)\eta^2\xi^{11}+\mathcal{O}(\xi^{10}).\eea
Therefore, since it was assumed that $\alpha_1\ne 0$ and $B_1 \ne 0$, the condition (\ref{CCN2}) requires that $A_{111}=A_{102}=A_{201}=0$ which is a contradiction because in this case all of the $A_{jk\ell}$'s would be identically zero. 
In the case that $B_1=0$, (\ref{CCN2}) is a polynomial in $\ln \xi$ with leading order term
\bea 0&=&\frac{K}{(\xi^2+\eta^2)^{10}\xi^4\eta^3} \ln^4 \xi+\mathcal{O}(\ln^3 \xi)\\
 K&=&192\alpha_1^6A_{111}\xi^{12}\eta^2-640\alpha_1^6A_{102}\xi^{11}\eta^3 -32(864 d_{-2}c_4A_{102}+\alpha^2A_{201})\alpha^4\xi^{13}\eta^3+\ldots
\eea
For simplicity, we give only the most relevant terms instead of the highest order ones in $K$. 
From these terms, it can be seen that the condition (\ref{CCN2}) leads to a contradiction since they would imply  $A_{201}$, $A_{102}$ and $A_{111}$ are all 0 and so every $A_{jk\ell}$ would vanish. 
Thus, there are no logarithmic singularities in the potential.

Next, we turn in particular to the case that $\alpha_1=\beta_1=0$. In this case, we proceed slightly differently because computationally, it is more difficult for MAPLE to compute coefficients of the expression (\ref{CCN2}) with respect to  $(\xi^2+\eta^2)^{k/2}$ without first simplifying the entire expression.  On the other hand, unlike in the previous section, we can solve (\ref{Eq1}-\ref{Eq3}) without much difficultly and replace the integrated forms of $G_1$ and $G_2$ into (\ref{Eq4}) to obtain the needed contradictions. 

In this case, we require the complete solutions for (\ref{cc}). Namely, $W_1$ must satisfy (\ref{W1m2}) with $c_3=c_5=0$ and $W_2$ must satisfy (\ref{W2m2}), both with $\alpha_1=\beta_1=0$. Additionally, the constants satisfy one of the following cases:
\bea\label{case11}\left\{  B_2=-B_1, A_{201}=A_{102}=0, c_0=c_1=d_1=c_6=0\right\}\\
\left\{B_2=-B_1, A_{201}=A_{111}=0, c_0=c_1=d_1=c_{-2}=0\right\}\\
\left\{ B_2=-B_1, A_{111}=0, A_{102}=A_{120}, c_1=d_1=c_{6}=c_{-2}=0\right\}\\
\left\{B_2=0,  A_{111}=0, A_{102}=A_{120}, c_0=c_1=d_1=c_{6}=c_{-2}=0\right\}\\
\left\{B_2=-B_1, c_0=c_1=d_1=c_{6}=c_{-2}=0\right\}.\eea
There is also a complex solution for the constants 
\be \left\{ B_2=-B_1, A_{201}=0, A_{111}=iA_{102}, d_1=ic_1, c_0=c_6=c_{-2}=0\right\}.\ee

To obtain the needed contradictions, we use the obtained sets of solutions for (\ref{cc}) to solve (\ref{Eq1}-\ref{Eq3}) for $G_1$ and $G_2$ and replace these solutions into (\ref{Eq4}). 
 For example, in the case identified in (\ref{case11}) the relevant $G's$ are 
 \bea  G_1&=&\left({\frac { -\left( -3\,B_{{1}}{\xi}^{2}+4\,{\xi}^{2}{\eta}^
{2}+{\eta}^{2}B_{{1}} \right) \alpha_{{2}}}{2\sqrt {{\xi}^{2}+B_{{1}}}
 \left( {\xi}^{2}+{\eta}^{2} \right) }}+{\frac {\xi\,\eta\, \left( -{
\eta}^{2}+2\,B_{{1}}+{\xi}^{2} \right) \beta_{{2}}}{\sqrt {{\eta}^{2}-
B_{{1}}} \left( {\xi}^{2}+{\eta}^{2} \right) }}\right.\nn&&
\left.-1/2\, \left( 2\,{\eta}
^{4}+{\eta}^{2}B_{{1}}+4\,{\xi}^{2}{\eta}^{2}-3\,B_{{1}}{\xi}^{2}
 \right) \xi\,c_{{4}}-{\frac { \left( {\eta}^{2}-{\xi}^{2}-B_{{1}}
 \right) c_{{-2}}}{{\eta}^{2}\xi}} \right) A_{{111}}\nn &&
 +\frac{ \eta(\xi^2+\eta^2)}{6}k_1+\eta k_2+\xi k_3\\
   G_2&=&\left( -{\frac {\xi\,\eta\, \left( -{\eta}^{2}+2\,B_{{1}}+{\xi}^{2}
 \right) \alpha_{{2}}}{\sqrt {{\xi}^{2}+B_{{1}}} \left( {\xi}^{2}+{
\eta}^{2} \right) }}-1/2\,{\frac { \left( -B_{{1}}{\xi}^{2}+4\,{\xi}^{
2}{\eta}^{2}+3\,{\eta}^{2}B_{{1}} \right) \beta_{{2}}}{\sqrt {{\eta}^{
2}-B_{{1}}} \left( {\xi}^{2}+{\eta}^{2} \right) }}\right. \nn&&
\left.+1/2\,\eta\, \left( 
2\,{\xi}^{4}-B_{{1}}{\xi}^{2}+4\,{\xi}^{2}{\eta}^{2}+3\,{\eta}^{2}B_{{
1}} \right) c_{{4}}+{\frac { \left( {\eta}^{2}-{\xi}^{2}-B_{{1}}
 \right) c_{{-2}}}{{\xi}^{2}\eta}} \right) A_{{111}}\nn&&
 -\frac{\xi(\xi^2+\eta^2)}{6}k_1-\eta k_3+\xi k_2
 ,\eea
where $k_1, k_2$ and $k_3$ are constants of integration. 
  These solutions for $G_1$ and $G_2$ when substituted into (\ref{Eq4}) give
  \be 0=T_1\sqrt{\xi^2+B_1}\sqrt{\eta^2-B_1} +T_2\sqrt{\xi^2+B_1} +T_3{\sqrt{\eta^2-B_1}}+T_4.\ee
  Again, the $T_i$'s are rational functions of $\xi$ and $\eta$, and can be obtained from the authors. 
 Solving these systems, we obtain $\beta_2=0$ and $\alpha_2=0$ or $A_{111=0}$, which gives a contradiction, since in these cases either all of the $A_{jk\ell}$ are zero or the potential reduces to a rational function.  By checking each case in this manner, we find that when $B_1$ is assumed to be non-zero the potentials reduce to rational functions.

Thus, we have shown by contradiction that the only possible potentials which satisfy both the linear (\ref{cc}) and nonlinear (\ref{ccnl1}-\ref{ccnl3}) compatibility conditions are rational functions of $\xi$ and $\eta.$

\section{Final list of superintegrable potentials}\label{Final}
Since there are no longer any irrational terms in the potential, it is a straightforward computation to find the admissible choices of constants which satisfy the linear compatibility condition (\ref{cc}), to use these choices to solve the linear partial differential equations  (\ref{Eq1}-\ref{Eq3}) for $G_1$ and $G_2$ and to solve the resulting algebraic system determined by the coefficients of (\ref{Eq4}).  
In this section, we exhibit the possible potentials which remain, i.e. those potentials which separate in parabolic coordinates and admit a third order integral of motion.
Remarkably, the only such potentials are second-order superintegrable. These are exactly the potentials which separate in parabolic coordinates as well as another orthogonal coordinate system in $E_2(\mathbb{R}).$ We also obtain a potential  in   $E(1,1)$ which is presented in Appendix \ref{lightcone}. 

It is interesting to note that, in addition to the systems which admit a single third-order integral, this method allows us to obtain potentials in the quantum case which admit more than one third-order integral. Furthermore, these additional potentials give proof that the quantum correction to (\ref{Eq4}), namely $\Phi,$ is not identically 0 for potentials which are superintegrable in both the classical and quantum cases. The quantity $\Phi$ only vanishes when, in addition, the appropriate choices of $A_{jkl}$ are assumed. For example, for potential $V_1$ below, the quantity $\Phi$ vanishes only when all of the $A_{jkl}$'s are assumed 0 except for $A_{012}.$

\subsection{Potentials which admit a single third-order integral}
\subsubsection{A deformation of the anisotropic oscillator potential: $V_1$}
The following potential 
\bea \label{V1}V_1&=&\left( {\eta}^{4}-{\xi}^{2}{\eta}^{2}+{\xi}^{4} \right) \alpha+\beta(\xi^2-\eta^2)+{\frac {\gamma
}{{\xi}^{2}{\eta}^{2}}}\\
&=&{\alpha}\left( 4\,{x}^{2}+{y}^{2} \right) +2\,\beta\,x+{\frac {
\gamma}{{y}^{2}}}\nn
&=&\alpha {r}^{2} \left( 3  \cos^2 \theta +1
 \right) +2\,\beta\,r\cos  \theta  +{\frac {\gamma}
{{r}^{2} \sin^2 \theta}}\nonumber
\eea
admits a third-order integral with all all $A_{jk\ell}=0$ except $A_{021}$ and functions
\bea G_{{1}}  =2\,{\frac {A_{{012}} \left( -2\,\alpha
\,{\xi}^{2}{\eta}^{6}+{\eta}^{4}\alpha\,{\xi}^{4}+{\eta}^{4}\beta\,{
\xi}^{2}+\gamma \right) }{\xi\,{\eta}^{2}}}
\nn
G_{{2}}  =-2\,{\frac {A_{{012}} \left( -2\,{\eta
}^{2}\alpha\,{\xi}^{6}+{\eta}^{4}\alpha\,{\xi}^{4}-{\eta}^{2}\beta\,{
\xi}^{4}+\gamma \right) }{\eta\,{\xi}^{2}}}.\nonumber
\eea

\subsubsection{A deformation of the Coulomb potential: $V_2$}
The following potential
\bea \label{V2} V_2&=& \frac{1}{\xi^2+\eta^2}\left(\frac{\alpha}{\xi^{2}}+
\frac{\beta}{{\eta}^{2} }+\gamma \right)\\
&=&\frac1{2\sqrt{x^2+y^2}}\left(\frac{\alpha}{{x+\sqrt{x^2+y^2}}}+\frac{\beta}{{x-\sqrt{x^2+y^2}}}+\gamma\right)\nn
&=& {\frac {\alpha}{2{r}^{2} \left( \cos  \theta  +1
 \right) }}-{\frac {\beta}{2{r}^{2} \left( \cos  \theta
  -1 \right) }}+{\frac {\gamma}{2r}}\nonumber
\eea
admits one third-order constant of motion with all $A_{jk\ell}=0$ except $A_{210}$ and functions
\bea G_{{1}} ={\frac {A_{{210}} \left( 2\,{
\eta}^{4}\gamma\,{\xi}^{2}+2\,{\eta}^{4}\alpha+4\,{\xi}^{2}\beta\,{
\eta}^{2}-{\eta}^{2}{\xi}^{2}{h}^{2}+2\,{\xi}^{4}\beta \right) }{4\xi\,
{\eta}^{2}}}\nn
G_{{2}} =-{\frac {A_{{210}} \left( 2\,{
\xi}^{4}\beta+2\,{\xi}^{4}\gamma\,{\eta}^{2}+4\,{\xi}^{2}\alpha\,{\eta
}^{2}-{\eta}^{2}{\xi}^{2}{h}^{2}+2\,{\eta}^{4}\alpha \right) }{4\eta\,{
\xi}^{2}}}. \nonumber
\eea

\subsubsection{A second deformation of the Coulomb potential: $V_3$}
The following potential
\bea\label{V3} V_3&=&{\frac {\alpha\,\xi+\beta\,\eta+\gamma}{{\xi}^{2}+{\eta}^{2}}}\\
&=&\frac{1}{2\sqrt{x^2+y^2}}\left(\alpha \sqrt{x+\sqrt{x^2+y^2}}+\beta \sqrt{\sqrt{x^2+y^2}-x}+\gamma\right)\nn
&=& \frac{1}{2r}\left(\alpha \sqrt{2} \cos \frac \theta 2+\beta \sqrt{2} \sin \frac \theta 2+\gamma\right)\nonumber
\eea
admits one third-order integral associated with $A_{102}$ (all the remaining $A_{jk\ell}=0$)
\bea G_{{1}}  =-\frac12A_{{102}} \left( -\beta\,{\xi}^{2}+\beta\,{\eta}^{2}+2\,\eta\,\alpha\,\xi+2\,\eta\,\gamma \right)\nn
G_{{2}}=\frac12 A_{{102}} \left( \alpha\,{\xi}^{
2}-\alpha\,{\eta}^{2}+2\,\xi\,\gamma+2\,\eta\,\xi\,\beta \right). \nonumber
\eea

\subsection{Potentials which admit more than one third-order integrals}
The following potentials are sub-cases of those from the previous section, in the quantum cases. They admit at least two third-order integrals. 

\subsubsection{$V_1$ subcases}
The following potential
 \bea V_{1, a}&=&\beta \left( {\xi}^{2}-{\eta}^{2} \right) +{\frac {{h}^{2}}{{\xi}^{2}
{\eta}^{2}}}\\
&=&2\beta x+\frac{\hbar^2}{y^2}\nonumber.\eea
admits three linearly independent third-order integrals associated with the constants $A_{012}, A_{003}$ and $A_{102}$ (the remaining $A_{jk\ell}$'s are 0). The functions $G_1$ and $G_2$ are
\bea  G_{{1}}  =-{\frac { \left( 3\,{h}^{2}{\eta
}^{2}-{h}^{2}{\xi}^{2}+2\,{\eta}^{4}{\xi}^{4}\beta \right) A_{{102}}}
{2\eta\,{\xi}^{2}}}+3\,{\frac {{h}^{2}A_{{003}}}{\eta\,{\xi}^{2}}}+2\,{
\frac { \left( {\eta}^{4}{\xi}^{2}\beta+{h}^{2} \right) A_{{012}}}{\xi
\,{\eta}^{2}}}\nn
 G_{{2}}  =-{\frac { \left( -3\,{h}^{2}{\xi
}^{2}+{h}^{2}{\eta}^{2}+2\,{\xi}^{4}{\eta}^{4}\beta \right) A_{{102}}
}{2\xi\,{\eta}^{2}}}+3\,{\frac {{h}^{2}A_{{003}}}{\xi\,{\eta}^{2}}}+2\,{
\frac { \left( {\xi}^{4}{\eta}^{2}\beta-{h}^{2} \right) A_{{012}}}{
\eta\,{\xi}^{2}}}.\nonumber\eea

 \subsubsection {$V_2$ subcases}
 The following potential 
\bea V_{2,a}&=&{\frac {\gamma}{{\xi}^{2}+{\eta}^{2}}}+{\frac {{h}^{2}}{{\xi}^{2}{\eta
}^{2}}}\\
&=& \frac{\gamma}{2r} +\frac{\hbar^2}{r^2\sin^2 \theta}.\nonumber\eea
admits two linearly independent third-order integrals associated with the constants $A_{300}$ and $A_{210}$ (the remaining $A_{jk\ell}$'s are 0). The functions $G_1$ and $G_2$ are
\bea  G_{{1}} =-{\frac {{h}^{2} \left( {\xi}^{2
}+{\eta}^{2} \right)  \left( 3\,{\xi}^{4}+2\,{\xi}^{2}{\eta}^{2}+3\,{
\eta}^{4} \right) A_{{300}}}{8\eta\,{\xi}^{2}}}
+{\frac { \left( 3
\,{\eta}^{2}{\xi}^{2}{h}^{2}+2\,{\eta}^{4}\gamma,{\xi}^{2}+2\,{\eta}^
{4}{h}^{2}+2\,{\xi}^{4}{h}^{2} \right) A_{{210}}}{4{\eta}^{2}\xi}}\nn
 G_{{2}}  ={\frac {{h}^{2} \left( {\xi}^{2}
+{\eta}^{2} \right)  \left( 3\,{\xi}^{4}+2\,{\xi}^{2}{\eta}^{2}+3\,{
\eta}^{4} \right) A_{{300}}}{8\xi\,{\eta}^{2}}}-{\frac { \left( 3
\,{\eta}^{2}{\xi}^{2}{h}^{2}+2\,{\xi}^{4}{h}^{2}+2\,{\eta}^{2}{\xi}^{4
}\gamma+2\,{\eta}^{4}{h}^{2} \right) A_{{210}}}{4\eta\,{\xi}^{2}}}.\nonumber
\eea
Another subcase of the potential $V_2$ which admits three linearly independent third-order integrals is given by 
\bea V_{2,b}&=&{\frac {1}{{\xi}^{2}+{\eta}^{2}}}\left(\gamma+{\frac {{h}^{2}}{{\xi}^{2}}}\right)\\
&=&\frac{\gamma}{2r}+\frac{\hbar^2}{2r^2(1+\cos \theta)}\nonumber
.\eea
The integrals are associated with coefficients $A_{300}, \, A_{210}$ and $A_{201}$ with the remaining $A_{jk\ell}=0$ and functions
\bea  G_{{1}}  &=&{\frac {{h}^{2}\eta\, \left( 3
\,{\eta}^{2}+2\,{\xi}^{2} \right)  \left( {\xi}^{2}+{\eta}^{2}
 \right) A_{{300}}}{8{\xi}^{2}}}
 +{\frac { \left( 2\,{h}^{2}{\eta}^
{2}+2\,{\eta}^{2}{\xi}^{2}\gamma-{h}^{2}{\xi}^{2} \right) A_{{210}}}{4
\xi}}\nn&&
-{\frac {A_{{201}}\eta\, \left(2\,{
\xi}^{4}\gamma  -3\,{h}^{2}{\eta}^{2}\right) }{4{\xi}^{2}}}\nn
 G_{{2}}  &=&{\frac {{h}^{2} \left( 3\,{\eta}
^{2}+2\,{\xi}^{2} \right)  \left( {\xi}^{2}+{\eta}^{2} \right) A_{{300
}}}{8\xi}}-{\frac {\eta\, \left( 2\,{\xi}^{4}\gamma+2\,{h}^{2}{
\eta}^{2}+3\,{h}^{2}{\xi}^{2} \right) A_{{210}}}{4{\xi}^{2}}}\nn&&+{
\frac {A_{{201}} \left( 2\,{\xi}^{4}\gamma-{h}^{2}{\eta}^{2} \right) }
{4\xi}}\nonumber.\eea

\subsection{Potentials admitting group symmetry}
The following potentials admit a Killing vector associated with group symmetry. 
\subsubsection{Rotational symmetry}
The Coulomb potential 
\be V_C={\frac {\alpha}{{\xi}^{2}+{\eta}^{2}}}=\frac{\alpha}{r}\ee
admits   four linearly independent third-order integrals associated with constants $A_{300}, A_{210}, A_{201}$ and $A_{102}$ as well as a Killing vector associated with $k_1$ in 
\bea  G_{{1}} = A_{{210}} \frac{\xi\, \left( 2\,\alpha\,{\eta}^{
2}-{h}^{2} \right)}{4}-\eta\,A_{{102}}\alpha-A_{{201}}\frac{\eta\, \left( 
2\,\alpha\,{\xi}^{2}+{h}^{2} \right)}4 -k_1\frac{\eta\,
 \left( {\xi}^{2}+{\eta}^{2} \right)}{2} \nn
  G_{{2}}  =- A_{{210}}\frac{\eta\, \left( 2\,\alpha\,{\xi}^
{2}-{h}^{2} \right)}{4}+A_{{102}}\xi\,\alpha+A_{{201}}\frac{\xi\, \left( 2
\,\alpha\,{\xi}^{2}-{h}^{2} \right)}{4} +k_1\frac{\xi\,
 \left( {\xi}^{2}+{\eta}^{2} \right)}2. \nonumber
 \eea
\subsubsection{Translation symmetry}
The first potential is 
\be V_{T,y}={\frac {\alpha}{{\xi}^{2}
{\eta}^{2}}}=\frac{\alpha}{y^2}.\ee
It admits 4 linearly independent constants of the motion associated with the constants $ A_{{012}},A_{{030}},A_{{111}},A_{{210}}$ and a Killing vector associated with $k_1$, with functions
\bea  G_{{1}}  ={\frac {\alpha\, \left( {\eta}^{
4}+{\xi}^{4} \right) A_{{210}}}{2\xi\,{\eta}^{2}}}-{\frac {\alpha\,
 \left( \eta-\xi \right)  \left( \eta+\xi \right) A_{{111}}}{\xi\,{
\eta}^{2}}}+2\,{\frac {A_{{012}}\alpha}{\xi\,{\eta}^{2}}}+k_1\,
\xi
\nn
 G_{{2}} =-{\frac {\alpha\, \left( {\eta}^
{4}+{\xi}^{4} \right) A_{{210}}}{2\eta\,{\xi}^{2}}}+{\frac {\alpha\,
 \left( \eta-\xi \right)  \left( \eta+\xi \right) A_{{111}}}{\eta\,{
\xi}^{2}}}-2\,{\frac {A_{{012}}\alpha}{\eta\,{\xi}^{2}}}-k_1\,
\eta.
\nonumber\eea

The potential 
\be V_{T,y, \hbar}={\frac {\hbar^2}{{\xi}^{2}
{\eta}^{2}}}=\frac{\hbar^2}{y^2}\ee
admits 8 linearly independent third-order constants of the motion associated with constants $ A_{{003}}, $ $ A_{{012}},$ $ A_{{030}},$ $ A_{{102}}, $ $A_{{111}}, $ $A_{{201}}, $ $A_{{
210}},$ and $A_{{300}}$ as well as a Killing vector associated with the constant $k_1$. The $G$'s are 
\bea  G_{{1}} ={\frac {{h}^{2} \left( 3\,{\eta}
^{4}-{\xi}^{4} \right) A_{{201}}}{4\eta\,{\xi}^{2}}}+{\frac {{h}^{
2} \left( {\eta}^{4}+{\xi}^{4} \right) A_{{210}}}{2\xi\,{\eta}^{2}}}-{\frac {{h}^{2} \left( 3\,{\eta}^{6}+5\,{\xi}^{2}{\eta}^{4}+5\,{\xi
}^{4}{\eta}^{2}+3\,{\xi}^{6} \right) A_{{300}}}{8\eta\,{\xi}^{2}}}
+\xi k_1
\nn
 G_{{2}}={\frac {{h}^{2} \left( 3\,{\xi
}^{4}-{\eta}^{4} \right) A_{{201}}}{4\xi\,{\eta}^{2}}}-{\frac {{h}
^{2} \left( {\eta}^{4}+{\xi}^{4} \right) A_{{210}}}{2\eta\,{\xi}^{2}}}+
{\frac {{h}^{2} \left( 3\,{\eta}^{6}+5\,{\xi}^{2}{\eta}^{4}+5\,{
\xi}^{4}{\eta}^{2}
+3\,{\xi}^{6} \right) A_{{300}}}{8\xi\,{\eta}^{2}}}-\eta k_1.\nonumber\eea

Finally, the potential 
\be V_{T,x}=\alpha(\xi^2-\eta^2)=2\alpha x\ee
admits four linearly independent third-order integrals associated with constants  $A_{003},$ $ A_{102},$ $ A_{021},$ $ A_{012}$ and a Killing vector associated with $k_1$, with 
\bea G_{{1}} =-{\eta}^{3}{\xi}^{2}A_{{102}}\alpha-2
\,{\eta}^{3}A_{021}\alpha+2\,{\eta}^{2}\xi\,A_{{012}}\alpha+2\,\eta\,{
\xi}^{2}A_{021}\alpha+\eta\,k_1
 \nn
G_{{2}} =-{\xi}^{3}A_{{102}}\alpha,{\eta}^{2}+
2\,{\xi}^{3}A_{021}\alpha+2\,\eta\,A_{{012}}{\xi}^{2}\alpha-2\,\xi\,A_{{
21}}\alpha ,{\eta}^{2}+\xi\,k_1.\nonumber
 \eea
 \section{Conclusions}
 The main conclusion of this article is a negative one: all third-order integrals for potentials separating in parabolic coordinates in $E_2(\mathbb{R})$ are reducible. The corresponding potentials allow separation of variables in at least two coordinate systems and are already known to be quadratically superintegrable \cite{FMSUW, WSUF}. Thus $V_1$ (\ref{V1}) is a deformed anisotropic harmonic oscillator, separable also in Cartesian coordinates. The potential $V_2$ (\ref{V2}) is a deformed Coulomb potential, separable also in polar coordinates. The potential $V_3$ (\ref{V3}) is a different deformation of the Coulomb potential, separable in two different parabolic coordinate systems (with the directrix of the parabolas either along the $x$ or the $y$ axis).  Since these potentials are second-order superintegrable, they are the same in classical and quantum mechanics. 
 
 This is in stark contrast with the results obtained in the case of potentials separating in Cartesian \cite{Gravel} or polar coordinates \cite{TW20101}. There the results were much richer, mainly because of the role played by the linear compatibility condition (\ref{cc}). 

Indeed, let us first consider a potential separating in Cartesian coordinates $V=W_1(x)+W_2(y).$ The linear compatibility conditions for this potential, written in Cartesian coordinates is given by 
\bea -F_{3}W_{1,xxx}+2(F_{2,y}-F_{3,x})W_{1,xx}-(3F_{1,yy}-2F_{2,xy}+F_{3,xx})W_{1,x}=\nn F_2W_{2,yyy}+(F_{2,y}-F_{3,x})W_{2,yy}+(F_{2,yy}-2F_{3,xy}+3F_{4,xx})W_{2y}\label{ccC}\eea
with 
\bea F_1&=&A_{{300}}{y}^{3}+A_{{210}}{y}^{2}-A_{{120}}y+A_{{030}}\nn
F_2&=&3\,A_{{300}}x{y}^{2}-2\,A_{{210}}xy+A_{{201}}{y}^{2}+A_{{120}}x-A_{{
111}}y+A_{{021}}\nn
F_3&=&-3\,A_{{300}}{x}^{2}y+A_{{210}}{x}^{2}-2\,A_{{201}}xy+A_{{111}}x-A_{{
102}}y+A_{{012}}\nn
F_4&=&A_{{300}}{x}^{3}+A_{{201}}{x}^{2}+A_{{102}}x+A_{{003}}.\nonumber \eea
Differentiating (\ref{ccC}) twice with respect to $x$ gives two linear ODEs for $W_1$:
\bea\left( 3\,A_{{300}}{x}^{2}+2\,A_{{201}}x+A
_{{102}} \right)W_1^{(5)}+\left( 36\,A_{{300}}x+12\,A_{{201}} \right) W_{1}^{(4)}+ 84\,A_{{300}}W_{1}^{(3)}
 =0\label{W11},\\
 \left( -A_{{111}}x-A_{{210}}{x}^{2}-A_{{012
}} \right) W_1^{(5)} +
 \left( -12\,A_{{210}}x-6\,A_{{111}} \right)W_1^{(4)} -28\,A_{{210}} W_1^{(3)}=0.\label{W12}\eea
Differentiating (\ref{ccC}) twice with respect to $y$ gives two linear ODEs for $W_2$:
\bea \left( 3\,A_{{300}}{y}^{2}-2\,A_{{210}}y+A_{{120}} \right) W_2^{(5)}+
 \left( 36\,A_{{300}}y-12\,A_{{210}} \right) W_2^{(4)}+  90\,A_{{300}}W_1^{(3)}=0\label{W21}\\
 \label{W22} \left( A_{{201}}{y}^{2} -A_{{111}}y+A_{{021}}\right) W_{2}^{(5)}+\left( 12\,A_{{201}}y-6\,A_{{111}} \right) W_2^{(4)}+ 30\,A_{{201}} W_2^{(3)} =0.\eea
 
 Interesting potentials in quantum mechanics, involving elliptic functions or Painlev\'e trascendents are obtained if either of these pairs of linear equations are satisfied trivially, i.e. if the coefficients in  (\ref{W11}-\ref{W12}) or (\ref{W21}-\ref{W22}) vanish identically. Both systems of equations (\ref{W11}-\ref{W22}) vanish identically if and only if  the only non-zero coefficients $A_{jk\ell}$ are $A_{030}$ and $A_{003}.$ The potential in this case is for instance \cite{Gravel} 
 \[V(x,y)=\hbar^2\left[\omega_1^2 P_I(\omega_1 x) +\omega^2_2P_I(\omega_1 y)\right],\]
 where $P_I$ is the first Painlev\'e transcendent\cite{Ince}. If, for example, (\ref{W21}-\ref{W22}) are satisfied trivially but  (\ref{W11}-\ref{W12}) are not, this leads to potentials of the form\cite{Gravel} 
 \[ V(x,y)=ay+(2\hbar^2b^2)^{1/3}\left[P_{II}'\left(-\left(\frac{4b}{\hbar^2}\right)^{1/3}x, k\right)+P_{II}^2\left(-\frac{4b}{\hbar^2}x,k\right)\right],\]
 where $P_{II}$ is the second Painlev\'e transcendent\cite{Ince}. 
 
 Now, let us consider potentials allowing separation of variables in polar coordinates\cite{TW20101} 
 \be V=R(r)+\frac{1}{r^2}S(\theta).\ee
 The linear compatibility condition reduces to 
 \bea\ba{rl} 0&= r^4 F_3R'''+ \left( 2{r}^{4} F_{{3,r}} -2{r}^{2}F_{{2,\theta}}+3r \left( 2
\,{r}^{2}F_{{3}} -F_{{1}}   \right)  \right) R'' \\ &
+\left({r}^{4}F_{{3,rr}}  +2\,{r}^{2}(3F_{{3,r}} +3F_{{
3}}-F
_{{2,r\theta }}) -4\,r F
_{{2, \theta}} +3 F_{{1, \theta\theta}} \right) R'\\ &
+\frac{1}{r^2}F_2S'''-\frac{1}{r^3}\left(2r^3F_{3,r}-2rF_{2,\theta}+6F_1\right)S''\\ &+\frac{1}{r^3}\left(3r^5F_{4,rr}+6r^4F_{4,r}-2r^3F_{3,r\theta}+3r^2F_{2,r}+r(F_{2,\theta\theta}-2F_2)-12F_{1,\theta}\right)S'\\ &-\frac{1}{r^3}\left(2r^4F_{3,rr}-12r^3F_{3,r}+4r^2(3F_{3,r}-F_{2,r\theta})+4rF_{2,\theta}+6F_{1,\theta\theta}+18F_1\right)S\ea \label{ccP},\eea
where here the $F_i$'s are given by
{
\bea \ba{rl} F_1&=A_1\cos 3\theta +A_2\sin 3\theta +A_3 \cos \theta +A_4\sin\theta\\ 
F_2&=\frac{-3A_1\sin 3\theta+3A_2\cos 3\theta -A_3\cos \theta +A_4\sin \theta}{r}+B_1\cos 2\theta +B_2\sin 2 \theta+B_0\\
F_3&= \frac{-3A_1\cos 3\theta-3A_2\sin 3\theta +A_3\cos \theta +A_4\sin \theta}{r^2}+\frac{-2B_1\sin 2\theta +2B_2\cos 2 \theta}{r}+C_1\cos\theta +C_2\sin \theta\\
F_4&=\frac{A_1\sin 3 \theta -A_2\cos 3\theta -A_3\sin \theta +A_4 \cos \theta}{r^3} -\frac{B_1\cos 2\theta+B_2\sin 2 \theta +B_0}{r^2}-\frac{C_1\sin\theta-C_2\cos\theta}{r}+D_0, \ea \nonumber\eea}
with 
\bea \ba{llll}A_1=\frac{A_{030}-A_{012}}{4}, &A_2=\frac{A_{021}-A_{003}}{4},& A_3=\frac{3A_{030}+A_{012}}{4},&\\
A_4=\frac{3A_{003}+A_{021}}{4}, & B_1=\frac{A_{120}-A_{102}}{2},&B_2=\frac{A_{111}}2, & B_0=\frac{A_{120}+A_{102}}2,\nn
C_1=A_{210}, & C_2=A_{201}, & D_0=A_{300}.& \ea\nonumber\eea
Equation (\ref{ccP}) is satisfied trivially if all the $A_{jk\ell}=0$ except $A_{300}$. This leads to the potential\cite{TW20101}
\[ V(r, \theta)=R(r)+\frac{\hbar^2}{r^2}P(\theta, t_2, t_3),\]
where $R(r)$ is arbitrary and  $P(\theta, t_2, t_3)$ is the Weierstrass elliptic function \cite{ByrdFriedman}. This potential allows a third-order integral, however it is algebraically related to the second-order one and the system is hence not superintegrable. 
If however we consider the subcase $R(r)=0$, Eq. (\ref{ccP}) simplifies. It being satisfied trivially allows further constants in the integral  to be nonzero, namely $A_{300}$, $A_{210}$ and $A_{201}$. This leads to a superintegrable potential expressed in terms of the sixth Painlev\'e transcendent\cite{Ince}, $P_{VI}(\sin\theta/2).$

In section \ref{general}, we have shown that the linear compatibility condition (\ref{cc}) is never satisfied trivially for a potential separating in parabolic coordinates. This rules out any dependence of $W_1$ and $W_2$ on elliptic functions or Painlev\'e transcendents. 

In addition to the real potentials presented in section \ref{Final}, we have recovered a complex one (see Appendix \ref{lightcone}) which is known to be superintegrable\cite{Eis}. It can be transformed into a real form; however, it will live on the pseudo-Euclidean plane $E(1,1)$ rather than the Euclidean one.

\begin{acknowledgments}
We thank F. Tremblay for helpful discussions. 
SP acknowledges a postdoctoral fellowship provided by the Laboratory of Mathematical Physics of the Centre de Recherches Math\'ematiques, Universit\'e de Montr\'eal. The research of PW is  partially supported by the National Sciences and Engineering Research Council (NSERC) of Canada.
\end{acknowledgments}

\appendix

\section{System of linear ODEs for $W_1$ and $W_2$}\label{eqs}
The linear compatibility condition (\ref{cc}), implies (by differentiation) two systems of 14 ODEs for $W_1$ and $W_2$, respectively. Those for $W_1$ are given below and those for $W_2$ can be obtained from these under the interchange $\xi\rightarrow \eta$ and $A_{jk\ell} \rightarrow (-1)^{j+k}A_{jk\ell}.$

{\small
\bea
 \label{eq1}  &&\ba{rl} 0=&A_{300} \left({\xi}^{2}W_1^{\left(17\right)}+37\xi\, W_1^{\left(16\right)}+323 W_1^{\left(15\right)}\right)\ea\\
 \label{eq2} &&\ba{rl} 0=& A_{{210}}\left(\xi\, W_1^{\left(17\right)}+18W_1^{\left(16\right)} \right)\ea\\
\label{eq3} &&\ba{rl} 0=&A_{210}\left( 11\,{\xi}^{3} W_1^{\left(17\right)}+503 {\xi}^{2}
 W_1^{\left(16\right)}+7145\,\xi W_1^{\left(15\right)}+
31360 W_1^{\left(14\right)}\right)\ea \\
  \label{eq4}&&\ba{rl} 0=&-9A_{300}\left({\xi}^{4}{W_{{1}}}^{(17)}+65{\xi}^{3}{W_{{1}}}^{(16)}+1485{\xi}^{2}{W_{{1}}}^{(15)}+14070\xi {W_{{1}}}^{(14)}+ 46410 {W_{{
1}}}^{(13)}\right)\\
&+(A_{{201}}{\xi}^{2}-2A_{{120}} )W_1^{(17)}
-33\xi\,A_{{201}} W_1^{(16)} -255A_{{201}}W_1^{(15)}.\ea \\
\label{eq5} && \ba{rl} 0=&-\frac{45}{4} A_{{300}}\left(
{\xi}^{6}{W_{{1}}}^{(17)} +93{\xi}^{5}{W_{{1}}}^{(16)}+3375{\xi}^{4}{W_{{1}}}^{(15)}+60900{\xi}^{3}{W_{{1}}}^{(14)}+573300{\xi}^{2}{W_{{1}}}^{(13)}\right.\\  &\left.  +2653560\xi {W_{{1}}}^{(12)}+
4684680\,{W_{{1}}}^{(11)}
 \right)\\
&
 - A_{{201}}\left( \frac {5{\xi}^{4}}2{W_{{1}}}^{(17)}+{\frac {
301{\xi}^{3}}{2}}\,{W_{{1}}}^{(16)}+{
\frac {6345{\xi}^{2}}{2}}\,{W_{{1}}}^{(15)}+{ {27615}\xi}{W_{{1}}
}^{(14)}+{ {83265}}\,{W_{{1}}}^{(13)} \right) \\
&- A_{{120}}\left( 6{\xi}^{2}{W_{{1}}}^{(17)}+{
 {169}}\xi {W_{{1}}}^{(16)}+{ {1109}}{W_{{1}}}^{(15)}
 \right) \\&
 + \left( {\xi}^{2}A_{{102}}+2A_{{021}} \right) {W
_{{1}}}^{17}+{ {33}}\xi\,A_{{102}}{W_{{1}}}^{(16)}+{ {255
}}A_{{102}}{W_{{1}}}^{(15)}.\ea\\
&& \label{eq6}  \ba{rl} 0=& \frac{-15}{2} A_{{300}}\left({\xi}^{8}{W
_{{1}}}^{(17)}+121{\xi}^{7}{W_{{1}}}^{(16)}+5993{\xi}^{6}{W_{{1}}}^
{(15)}+157962{\xi}^{5}{W_{{1}}}^{(14)}+2410590{\xi}^{4}{W_{{1}}}^{(13)}\right.\\ &\left. +21676200{\xi}^{3}{W_{{1}}}^{(12)}+111351240{\xi}^{2}{W_{{1}}}^{(11)}+296215920\xi {W_{{1}}}^{
(10)}+309188880\,{W_{{1}}}^{(9)}
 \right) \\&
-A_{{201}} \left( \frac{5{\xi}^{6}}2{W_{{1}}}^{(17)}+{\frac {437{\xi}^{5}}{2}}{W_{{1}}}^{(16)}+{\frac {14831{\xi}^{4}}{2}}\,{W_{{1}}}^{(15)} +124418{\xi}^{3}{W_{{1}}}^{(14)}\right.\\&\left.+1081626{\xi}^{2}{W
_{{1}}}^{(13)}+ 
4585308\xi{W_{{1}}}^{(12)}+7339332\,{W_{{1}}}^{(11)}\right) \\
&-A_{{120}} \left( 7{\xi}^{4}{W_{{1}}}^{(17)}+388{\xi}^{3}{W_{{1}}}^{(16)}+7464{\xi}^{2}{W_{{1}}
}^{15}+58632\xi {W_{{1}}}^{(14)}+157248 {W_{{1}}}^{(13)} \right) \\
&
+ 2A_{{102}} \left( 
{\xi}^{4}{W_{{1}}}^{(17)}+58{\xi}^{3} {W_{{1}}}^
{(16)}+1170{\xi}^{2}{W_{{1}}}^{(15)}+9660\xi {W_{{1}}}^{(14)}+27300{W_{{1}}}^{(13)}
 \right)\\ &
+A_{{021}} \left(
{\xi}^{2}{W_{{1}}}^{(17)}-11\xi{W_{{1}}}^{(16)} -409{W_{{1}}}^{(15)} \right) \\ 
 & +A_{003}\left(3{\xi}^{2} {W_{{1}}}^{(17)}+111\xi {W_{{1}}}^{(16)}+969\,{W_{{1}}}^{(15)}\right).\ea\\
 \label{eq10} && \ba{rl} 0=&A_{{210}}\left( {
\frac {25{\xi}^{5}}{4}}\,{W_{{1}}}^{(17)}+{
\frac {1839{\xi}^{4}}{4}}\,{W_{{1}}}^{(16)}+{\frac {50451{\xi}^{3}}{4}}\,{W_{{1}}}^{(15)}+{\frac {320901{\xi}^{2}}{2}}\,{
W_{{1}}}^{14}+{\frac {1881243\xi}{2}}\,{W_{{1}}}^{(13)}\right.\\ &\quad \left.+2018016\,{W_{{1}}}^{(12)} \right) \\&+ \left( 3
\,\xi\,A_{{030}}+1/2\,{\xi}^{3}A_{{111}}-2\,\xi\,A_{{012}} \right) {W_{{
1}}}^{17}+ \left( 54\,A_{{030}}-36\,A_{{012}}+{\frac {51}{2}}\,{\xi}^{2}
A_{{111}} \right) {W_{{1}}}^{(16)}\\
&+{\frac {813}{2}}\,{W_{{1}}}^{(15)}\xi\,
A_{{111}}+2016\,{W_{{1}}}^{(14)}A_{{111}}.\ea\\
&& \label{eq12} \ba{rl} 0=&\frac{1}{4}\,A_{{210}}\left( 25\,{W_{{1}}}^{(17)}{\xi}^{5}+1839\,{W_{{1}}}^
{(16)}{\xi}^{4}+50451\,{W_{{1}}}^{(15)}{\xi}^{3}+641802\,{W_{{1}}}^{(14)}{\xi}^
{2}+3762486\,{W_{{1}}}^{{(13)}}\xi\right.\\&\quad\left. +8072064\,{W_1}^{(12)} \right)
\\
&+ \frac{1}{2}\,A_{{111}}\left( {W_{{1}}}^{(17)}{\xi}^{3}+51\,{W_{{1}}}^{(16)}{
\xi}^{2}+813\,{W_{{1}}}^{(15)}\xi+4032\,{W_1}^{(14)} \right)
\\ 
&+3\,A_{{030}}\left( {W_{{1}}}^{(17)}\xi+18\,{W_1}^{(16)} \right)-2\,A_{{012}} \left({ W_{{1}}}^{(17)}\xi+18\,{W_1}^{(16)} \right) .\ea\\
&&\label{eq13} \ba{rl} 0=&\frac{1}{2}\, A_{{210}} \left( 15\,{W_{{1}}}^{(17)}{\xi}^{7}+1523\,{W_{{1}}}^
{(16)}{\xi}^{6}+61761\,{W_{{1}}}^{(15)}{\xi}^{5}+1289820\,{W_{{1}}}^{(14)}{\xi}
^{4}+14889420\,{W_{{1}}}^{(13)}{\xi}^{3}\right.\\&\quad\left. +94316040\,{W_{{1}}}^{(12)}{\xi}^{2}
+300900600\,{W_{{1}}}^{(11)}\xi+369008640\,{W_1}^{(10)} \right)
 \\
&+ 2\,A_{{111}}\left( {W_{{1}}}^{(17)}{\xi}^{5}+76\,{W_{{1}}}^{(16)}{\xi}
^{4}+2156\,{W_{{1}}}^{(15)}{\xi}^{3}+28392\,{W_{{1}}}^{(14)}{\xi}^{2}+172536
\,{W_{{1}}}^{(13}\xi\right.\\&\quad\left. +384384\,{W_1}^{(12)} \right)
\\ &
-A_{{012}}\left( 5\,{W_{{1}}}^{(17)}{\xi}^{3}+173\,{W_{{1}}}^{(16)}{
\xi}^{2}+1475\,{W_{{1}}}^{(15)}\xi+1024\,{W_1}^{(14)} \right) \\&
3\,A_{030}\left( 3\,{W_{{1}}}^{(17)}{\xi}^{3}+115\,{W_{{1}}}^{(16)}{
\xi}^{2}+1249\,{W_{{1}}}^{(15)}\xi+3392\,{W_1}^{(14)} \right) 
. \ea   \\ 
&& \label{eq9} \ba{rl} 0=&
-\frac{9}{4}\,A_{{300}}\left( {W_{{1}}}^{(17)}{\xi}^{12}+177\,{W_{{1}}}^{(16)
}{\xi}^{11}+13413\,{W_{{1}}}^{(15)}{\xi}^{10}+572670\,{W_{{1}}}^{(14)}{\xi}
^{9}+15257970\,{W_{{1}}}^{(13)}{\xi}^{8}\right.\\&\quad\left. +265552560\,{W_{{1}}}^{(12)}{\xi}^{7
}+3072429360\,{W_{{1}}}^{(11)}{\xi}^{6}+23597814240\,{W_{{1}}}^{(10)}{\xi}^{
5}\right.\\ &\quad \left.+118118800800\,{W_{{1}}}^{(9)}{\xi}^{4} +370767196800\,{W_{{1}}}^{(8)}{
\xi}^{3}+681080400000\,{W_{{1}}}^{(7)}{\xi}^{2}\right.\\ &\quad \left.+642939897600\,{W_{{1}}}^{(6)}\xi
+228843014400\,{W_1}^{(5)} \right) 
\\
&-\frac{1}{4}\, A_{{201}}\left( 5\,{W_{{1}}}^{(17)}{\xi}^{10}+709\,{W_{{1}}}^
{(16)}{\xi}^{9}+42051\,{W_{{1}}}^{(15)}{\xi}^{8}+1365672\,{W_{{1}}}^{(14)}{\xi}
^{7}\right.\\ &\quad \left.+26708136\,{W_{{1}}}^{(13)}{\xi}^{6} +325909584\,{W_{{1}}}^{(12)}{\xi}^{5
}+2487204720\,{W_{{1}}}^{(11)}{\xi}^{4}\right.\\ &\quad \left.+11568997440\,{W_{{1}}}^{(10)}{\xi}^{
3}+30849698880\,{W_{{1}}}^{(9)}{\xi}^{2}\right.\\ &\quad \left. +41565363840\,{W_{{1}}}^{(8)}\xi+
20704844160\,{W_1}^{(7)} \right)
\\
&-\frac{1}{2}\,A_{{120}}\left( 9\,{W_{{1}}}^{(17)}{\xi}^{8}+976\,{W_{{1}}}^{(16)
}{\xi}^{7}+42728\,{W_{{1}}}^{(15)}{\xi}^{6}+978432\,{W_{{1}}}^{(14)}{\xi}^{5
}+12689040\,{W_{{1}}}^{(13)}{\xi}^{4}\right.\\&\quad\left. +94174080\,{W_{{1}}}^{(12)}{\xi}^{3}+
383423040\,{W_{{1}}}^{(11)}{\xi}^{2}+761080320\,{W_{{1}}}^{(10)}\xi+536215680\,{W_1}^{(9)}
 \right)
 \\
&+ 2\, A_{{102}
}\left( {W_{{1}}}^{(17)}{\xi}^{8}+108\,{W_{{1}}}^{(16)}{\xi}
^{7}+4704\,{W_{{1}}}^{(15)}{\xi}^{6}+107016\,{W_{{1}}}^{(14)}{\xi}^{5}+
1375920\,{W_{{1}}}^{(13)}{\xi}^{4}\right.\\&\quad\left. +10090080\,{W_{{1}}}^{(12)}{\xi}^{3}+
40360320\,{W_{{1}}}^{(11)}{\xi}^{2}+77837760\,{W_{{1}}}^{(10)}\xi+51891840\,{W_1}^{(9)}
 \right)
\\
&- A_{{021}}\left( 5\,{W_{{1}}}^{(17)}{\xi}^{6}+393\,{W_{{1}}}^{(16)}{
\xi}^{5}+11511\,{W_{{1}}}^{}{\xi}^{(15)}+157140\,{W_{{1}}}^{(14)}{\xi}^{3}+
1016820\,{W_{{1}}}^{(13)}{\xi}^{2}\right.\\&\quad\left. +2784600\,{W_{{1}}}^{(12)}\xi+2325960\,{W_1}^{(11)} \right)
\\
&+9\,A_{{003}}\left( {W_{{1}}}^{(17)}{\xi}^{6}+77\,{W_{{1}}}^{(16)}{\xi}
^{5}+2223\,{W_{{1}}}^{(15)}{\xi}^{4}+30180\,{W_{{1}}}^{(14)}{\xi}^{3}+196980
\,{W_{{1}}}^{(13)}{\xi}^{2}\right.\\&\quad\left. +556920\,{W_{{1}}}^{(12)}\xi+491400\,{W_1}^{(11)} \right)
 .\ea\\ 
&&  \label{eq7} \ba{rl} 
0=&-\frac{45}{8} A_{{300}}\left({\xi}^{10}{W_{{1}}}^{(17)}+149{\xi}^{9}{W_
{{1}}}^{(16)}+9339{\xi}^{8}{W_{{1}}}^{(15)}+322728{\xi}^{7}{W_{{1}}}^{(14)}\right.\\&\quad \left.+6772584{\xi}^{6}{W_{{1}}}^{(13)}+89618256{\xi}^{5}{W_{{1}}}^{(12)}+751710960{\xi}^{4}{W_{{1}}}^{(11)}+3912068160{\xi}^{3}{W_{{1}
}}^{(10)}\right.\\&\quad \left.+11961069120{\xi}^{2}{W_{{1}}}^{(9)}+19148088960\xi{W
_{{1}}}^{(8)}+11987015040\,{W_{{1}
}}^{(7)}
 \right)\\
 &-A_{{201}} \left({\frac{5{\xi}^{8}}{2}{W_{{1}}}^{(17)}+ \frac {573{\xi}^{7}}{2}}\,{W_{{1}}}^{(16)}+{\frac {26733{\xi}^{6}}{2}}\,{W_{{1}}}^{(15)}+ 329721{\xi}^{5}{W_{{1}}}^{(14)}+4672395{\xi}^{4}{W_{{1}}}^{(13)}  \right. \\ &\quad\left.+38640420{\xi}^{3}{W_{{1}}}^{(12)}+180360180{\xi}^{2}{W_{{1}}}^{(11)}+429188760\xi {W_{{1}}}^{(10)}+392432040\,{W_{{1}}}^{(9)} \right) 
\\&-A_{{120}} \left( 8{\xi}^{6}{W_{{1}}}^{(17)}+657{\xi}^{5}{W_{{1}}}^{(16)}+20769{\xi}^{4}{W_{{1}}}^{(15)}+320964{\xi}^{3}{W
_{{1}}}^{(14)}+2532348{\xi}^
{2}{W_{{1}}}^{(13)}\right.\\ &\quad \left. +9546264\xi {W_{{1
}}}^{(12)}+13189176{W_{{1}}}^{(11)} \right) \\ &
+ 3A_{{102}}\left({\xi}^{6}{W_{{1}}}^{(17)}+83{\xi}^{5}{W_{{1}}}^{(16)}+2653{\xi}^{4}{W_{{1}}}^{(15)}+41468{\xi}^{3}{W_{{1}}}^{(14)} \right. \\ &\quad \left.+330876{\xi}^{2} {W_{{1}}}^{
(13)}+1260168\xi {W_{{1}}}^{(12)}+1753752{W_{{1}}}^{(11)}\right) \\ 
&-3 A_{{021}}\left( {\xi}^{4}{W_{{1}}}^{(17)}+71{\xi}^{3}{W_{{1}}}^{(16)}+1583{\xi}^{2}{W_{{1}}
}^{(15)}+12858\xi{W_{{1}}}^{(14)} +29022{W_{{1}}}^{(13)}
 \right) \\
&+9A_{{003}} \left(
{\xi}^{4}{W_{{1}}}^{(17)}+57{\xi}^{3}{W_{{1}}}^{(16)}+ 1093{\xi}^{2}{W_{{1}}}^{(15)}
+8070\xi{W_{{1}
}}^{(14)} +17850{W_{{1}}}^{(13)} \right) .\ea \\ &&\label{eq8}  \ba{rl}0=& \frac{-8}3A_{300}\left({\xi}^{14}{W_{{1}}}^{(17)}+205{\xi}
^{13}{W_{{1}}}^{(16)}+18215{\xi}^{(12)}{W_{{1}}}^{(15)}+925260{\xi}^{11}{W_{{1}}}^{(14)}\right.\\ &\quad \left.+29849820{\xi}^{10}{W_{{1}}}^{(13)}+642762120{\xi}^{9}{W_{{1}}}^{(12)} +9454044600{\xi}^{8}{W_{{1}}}^{(11)}+95610715200{\xi
}^{7}{W_{{1}}}^{10}\right.\\ &\quad \left. +660712852800{\xi}^{6}{W_{{1}}}^{(9)} +3062396937600{\xi}^{5}{W_{{1}}}^{(8)}+9210023222400{\xi}^{4}{W_{{1}}}^{(7)} \right.\\ &\quad \left.+17036091072000{\xi}^{3}{W_{{1}}}^{(6)}+17762576832000{\xi}^{2}{W_{{1}}}^{(5)}\right.\\ &\quad \left. +8935774848000\xi {W_{{1}}}^{(4)}
+1525620096000{W_{{1}}}^
{(3)} \right)\\
&-\frac{1}{4} A_{{201}} \left( {W_{{1}}}^{(17)}{\xi}^{12}+169\,{W_{{1}}}^{(16)
}{\xi}^{11}+12157\,{W_{{1}}}^{(15)}{\xi}^{10}+489230\,{W_{{1}}}^{(14)}{\xi}
^{9}+12178530\,{W_{{1}}}^{(13)}{\xi}^{8}\right.\\&\quad\left. +195839280\,{W_{{1}}}^{(12)}{\xi}^{7
}+2063421360\,{W_{{1}}}^{(11)}{\xi}^{6}+14153499360\,{W_{{1}}}^{(10)}{\xi}^{
5}\right.\\ &\quad \left.+61556695200\,{W_{{1}}}^{(9)}{\xi}^{4} +161124163200\,{W_{{1}}}^{(8)}{\xi
}^{3}+230659228800\,{W_{{1}}}^{(7)}{\xi}^{2}\right.\\ &\quad \left.+148929580800\,{W_{{1}}}^{(6)}\xi+
25427001600\,{W_1}^{(5)}
\right)\\
& -\frac{1}{2}\, A_{{120}}\left( 2\,{W_{{1}}}^{(17)}{\xi}^{10}+269\,{W_{{1}}}^
{(16)}{\xi}^{9}+15009\,{W_{{1}}}^{(15)}{\xi}^{8}+453768\,{W_{{1}}}^{(14)}{\xi}^
{7}+8148504\,{W_{{1}}}^{(13)}{\xi}^{6}\right.\\&\quad\left. +89618256\,{W_{{1}}}^{(12)}{\xi}^{5}+
600359760\,{W_{{1}}}^{(11)}{\xi}^{4}+2355312960\,{W_{{1}}}^{(10)}{\xi}^{3}+
4955670720\,{W_{{1}}}^{(9)}{\xi}^{2}\right.\\&\quad\left. +4618373760\,{W_{{1}}}^{(8)}\xi+1089728640\,{W_1}^{(7)}
 \right)
\\
& +\frac{1}{2}\,  A_{{102}}\left( {W_{{1}}}^{(17)}{\xi}^{10}+133\,{W_{{1}}}^{(16)}{
\xi}^{9}+7323\,{W_{{1}}}^{(15)}{\xi}^{8}+217896\,{W_{{1}}}^{(14)}{\xi}^{7}+
3837288\,{W_{{1}}}^{(13)}{\xi}^{6}\right.\\&\quad\left. +41185872\,{W_{{1}}}^{(12)}{\xi}^{5}+
267387120\,{W_{{1}}}^{(11)}{\xi}^{4}+1006125120\,{W_{{1}}}^{(10)}{\xi}^{3}+
1997835840\,{W_{{1}}}^{(9)}{\xi}^{2}\right.\\&\quad\left. +1712430720\,{W_{{1}}}^{(8)}\xi+363242880\,{W_1}^{(7)}
 \right)
\\
&- A_{{021}} \left( 2\,{W_{{1}}}^{(17)}{\xi}^{8}+191\,{W_{{1}}}^{(16)}{\xi
}^{7}+7171\,{W_{{1}}}^{(15)}{\xi}^{6}+135846\,{W_{{1}}}^{(14)}{\xi}^{5}+
1385370\,{W_{{1}}}^{(13)}{\xi}^{4}\right.\\&\quad\left. +7502040\,{W_{{1}}}^{(12)}{\xi}^{3}+
19819800\,{W_{{1}}}^{(11)}{\xi}^{2}+20900880\,{W_{{1}}}^{(10)}\xi+5045040\,{W_1}^{(9)} \right)
 \\ &+
 3\,A_{{00 3}}\left( {W_{{1}}}^{(17)}{\xi}^{8}+97\,{W_{{1}}}^{(16)}{\xi}^
{7}+3713\,{W_{{1}}}^{(15)}{\xi}^{6}+72090\,{W_{{1}}}^{(14)}{\xi}^{5}+759150
\,{W_{{1}}}^{(13)}{\xi}^{4}\right.\\&\quad\left. +4291560\,{W_{{1}}}^{(12)}{\xi}^{3}+12022920\,{W_
{{1}}}^{(11)}{\xi}^{2}+13693680\,{W_{{1}}}^{(10)}\xi+3603600\,{W_1}^{(9)} \right) 
 .\ea\\ && \label{eq14} \ba{rl} 0=&  \frac{1}{2}\, A_{{210}}\left( 10\,{W_{{1}}}^{(17)}{\xi}^{9}+1297\,{W_{{1}}}^{
(16)}{\xi}^{8}+69655\,{W_{{1}}}^{(15)}{\xi}^{7}+2022230\,{W_{{1}}}^{(14)}{\xi}^
{6}+34761090\,{W_{{1}}}^{(13)}{\xi}^{5}\right.\\&\quad\left.+364236600\,{W_{{1}}}^{(12)}{\xi}^{4}
+2306424120\,{W_{{1}}}^{(11)}{\xi}^{3}+8421613200\,{W_{{1}}}^{(10)}{\xi}^{2}\right.\\ &\quad \left.
+15881065200\,{W_{{1}}}^{(9)}\xi +11589177600\,{W_1}^{(8)} \right) 
\\
&+ 3\,A_{{
111}} \left( {W_{{1}}}^{(17)}{\xi}^{7}+101\,{W_{{1}}}^{(16)}{\xi
}^{6}+4067\,{W_{{1}}}^{(15)}{\xi}^{5}+84140\,{W_{{1}}}^{(14)}{\xi}^{4}+
959140\,{W_{{1}}}^{(13)}{\xi}^{3}\right.\\&\quad\left. +5973240\,{W_{{1}}}^{(12)}{\xi}^{2}+
18618600\,{W_{{1}}}^{(11)}\xi+22102080\,{W_1}^{(10)} \right)
\\ & +9\,A_{{030}}\left( {W_{{1}}}^{(17)}{\xi}^{5}+59\,{W_{{1}}}^{(16)}{\xi}
^{4}+1227\,{W_{{1}}}^{(15)}{\xi}^{3}+10946\,{W_{{1}}}^{(14)}{\xi}^{2}+40838
\,{W_{{1}}}^{(13)}\xi\right.\\&\quad\left. +53872\,{W_1}^{(12)} \right)\\
 & -3\,A_{{012}}\left( {W_{{1}}}^{(17)}{\xi}^{5}+41\,{W_{{1}}}^{(16)}{\xi
}^{4}+273\,{W_{{1}}}^{(15)}{\xi}^{3}-6146\,{W_{{1}}}^{(14)}{\xi}^{2}-78638\,
{W_{{1}}}^{(13)}\xi\right.\\&\quad\left. -206752\,{W_1}^{(12)} \right) 
.\ea\\
 && \label{eq11}  \ba{rl} 0=&\frac{1}{4}\, A_{{210}}\left( 7\,{W_{{1}}}^{(17)}{\xi}^{11}+1107\,{W_{{1}}}^
{(16)}{\xi}^{10}+74133\,{W_{{1}}}^{(15)}{\xi}^{9}+2760408\,{W_{{1}}}^{(14)}{
\xi}^{8}\right.\\ &\quad \left. +63115416\,{W_{{1}}}^{(13)}{\xi}^{7} +923792688\,{W_{{1}}}^{(12)}{\xi
}^{6}+8757180432\,{W_{{1}}}^{(11)}{\xi}^{5}+53215081920\,{W_{{1}}}^{(10)}{
\xi}^{4}\right.\\ &\quad \left.+200587907520\,{W_{{1}}}^{(9)}{\xi}^{3} +439575776640\,{W_{{1}}}^
{(8)}{\xi}^{2}+493647073920\,{W_{{1}}}^{(7)}\xi\right.\\ &\quad \left.+209227898880\,{W_1}^{(6)} \right)
\\
& +  2A_{{111
}}\left( {W_{{1}}}^{(17)}{\xi}^{9}+126\,{W_{{1}}}^{(16)}{\xi}
^{8}+6546\,{W_{{1}}}^{(15)}{\xi}^{7}+182868\,{W_{{1}}}^{(14)}{\xi}^{6}+
3004092\,{W_{{1}}}^{(13)}{\xi}^{5}\right.\\&\quad\left. +29811600\,{W_{{1}}}^{(12)}{\xi}^{4}+
176576400\,{W_{{1}}}^{(11)}{\xi}^{3}+592431840\,{W_{{1}}}^{(10)}{\xi}^{2}+
998917920\,{W_{{1}}}^{(9)}\xi\right.\\&\quad\left. +622702080\,{W_1}^{(8)} \right)
\\ 
&+  3\,A_{{030}}\left( {W_{{1}}}^{(17)}{\xi}^{7}+81\,{W_{{1}}}^{(16)}{\xi}
^{6}+2535\,{W_{{1}}}^{(15)}{\xi}^{5}+39444\,{W_{{1}}}^{(14)}{\xi}^{4}+329364
\,{W_{{1}}}^{(13)}{\xi}^{3}\right.\\&\quad\left. +1500408\,{W_{{1}}}^{(12)}{\xi}^{2}+3583944\,{W_{{
1}}}^{(11)}\xi+3459456\,{W_1}^{(10)} \right)
\\ 
&+ A_{{012}}\left( {W_{{1}}}^{(17)}{\xi}^{7}+129\,{W_{{1}}}^{(16)}{\xi}^{
6}+6027\,{W_{{1}}}^{(15)}{\xi}^{5}+134484\,{W_{{1}}}^{(14)}{\xi}^{4}+1544004
\,{W_{{1}}}^{(13)}{\xi}^{3}\right.\\&\quad\left. +8969688\,{W_{{1}}}^{(12)}{\xi}^{2}+23633064\,{W_{
{1}}}^{(11)}\xi+20756736\,{W_1}^{(10)} \right)
 .\ea \eea
}
\section{Complex potentials and superintegrable systems in $E(1,1)$ admitting separation in parabolic coordinates}\label{lightcone}
As a byproduct of this study, we have obtained a complex potential in $E_2$ that can be viewed as real potential in the pseudo-Euclidean place $E(1,1)$. This is the only complex potential in Ref. \onlinecite{Eis} which is separable in parabolic coordinates and  does not admit a Killing vector. We also give the subcase of this potential which admits a Killing vector and its third-order integrals.

\subsection{System admitting a single third-order integral:$V_4$}
The following potential 
\bea V_4&=&-\alpha(\xi^2-\eta^2)+\beta {\frac {
 \left({\xi}^{2}+i\xi\,\eta  -{\eta}^{2}\right) }{\xi+i\eta}}+{
\frac {\gamma}{\xi+i\eta}},\\
&=&2\,\alpha\,x-{\frac {\beta\, \left( 2
\,x+iy \right) }{\sqrt {x+iy}}} -{\frac {\gamma}{\sqrt {x+iy}}}\nn
&=& 2\alpha r\cos \left( \theta \right) -\frac{r\left( 2\,
\cos  \theta  +i\sin \theta  \right) \beta+\gamma}{\sqrt {r \left(\cos  \theta +i\sin
\theta \right) }}\nonumber
\eea
admits one third-order integral with $A_{3}=-A_{021}, $ $A_{012}=2iA_{021}$ and the remained $A_{jk\ell}=0$. The functions $G_1$ and $G_2$ are given by
\bea G_{{1}} =i \left( -2\,i\eta\,\alpha\,{\xi}^{2}
+4\,\xi\,\alpha\,{\eta}^{2}-2\,i\eta\,\xi\,\beta+2\,i\alpha\,{\eta}^{3
}+{\eta}^{2}\beta+\gamma \right) A_{021}\nn
G_{{2}} = \left( -2\,\xi\,\alpha\,{\eta}^{2}+4
\,i\eta\,\alpha\,{\xi}^{2}+2\,i\eta\,\xi\,\beta+2\,{\xi}^{3}\alpha+
\beta\,{\xi}^{2}-\gamma \right) A_{021}.\nonumber
\eea
Putting $iy=t$, we obtain a superintegrable system in $E(1,1);$ namely,
\be H=\frac{1}2(p_x^2-p_t^2) +2\alpha x-{\frac {\beta\, \left( 2
\,x+t \right) }{\sqrt {x+t}}} -{\frac {\gamma}{\sqrt {x+t}}}.\ee

\subsection{System admitting group symmetry}
A special case of the previous potential admits a Killing vector. The potential is:
\be V_{L}=\frac {\gamma}{\xi+i\eta }.\ee
The third-order integral is defined by the following non-zero $A_{jk\ell}$, along with an arbitrary constant $k_1$ which is the coefficient of the Killing vector
\[ A_{003} = A_{030}=\frac{A_{021}+2iA_{012}}{3},   A_{120} = -iA_{111}+A_{102} , A_{102}, A_{111}, A_{021}, A_{012},\]
\bea  G_{{1}}=\gamma A_{{012}}+i\gamma\,A_{021} 
+i\gamma \left( \xi +i\eta \right) ^{2}A_{{102}}-\frac{\gamma}2
 \left( \xi^2-\eta^2\right) A_{{111}}+ k_1
\, \left( \xi+i\eta\right) \nn
 G_{{2}} =i\gamma A_{{012}}+\gamma A_{021}+\frac{\gamma}{2} \left( \xi+i\eta \right) ^{2}A_{{102}}-\frac{i\gamma}{2}
 \left( \xi^2-\eta^2 \right) A_{{111}}+i k_1
 \left( \xi+i\eta \right) \nonumber.\eea

\end{document}